\pdfoutput = 1
\documentclass[conference]{IEEEtran}
\ifCLASSINFOpdf
\else
\fi

\usepackage{amsmath,amssymb,amsfonts}
\usepackage[utf8]{inputenc}
\usepackage{graphicx}

\usepackage{xspace,paralist}
\usepackage{amsthm}
\usepackage{tikz}
\let\labelindent\relax
\usepackage{enumitem}
\usepackage{multirow}
\usepackage{makecell}
\usepackage{tasks}

\usepackage[sort, compress]{cite}

\usepackage{subcaption}
\usepackage{caption}
\usepackage{tabularx}
\usepackage{multicol}
\usepackage{subfloat}
\usepackage{algorithm}
\usepackage{algcompatible}
\usepackage[noend]{algpseudocode}
\usepackage{filecontents}
\usepackage{mathtools}
\usepackage{comment}
\usepackage{url}

\usepackage{hyperref}
\hypersetup{colorlinks,urlcolor=blue}

\makeatletter
\def\Bstate{\State\hskip-\ALG@thistlm}
\makeatother



\newcommand{\xmark}{\ding{55}}

\newcommand{\paragraphb}[1]{\vspace{0.03in} \noindent{\bf #1} }
\newcommand{\paragraphe}[1]{\vspace{0.03in} \noindent{\em #1} }
\newcommand{\paragraphbe}[1]{\vspace{0.03in} \noindent{\bf \em #1} }

\newcommand{\grumbler}[2]{\textcolor{red}{\bf #1: #2}}

\newcommand{\oogrumbler}[2]{\textcolor{blue}{\bf #1: #2}}

\newcommand{\amir}[1]{\grumbler{Amir}{#1}}

\newcommand{\ali}[1]{\oogrumbler{Ali}{#1}}

\newcommand{\StatexIndent}[1][3]{%
  \setlength\@tempdima{\algorithmicindent}%
  \Statex\hskip\dimexpr#1\@tempdima\relax}

\usepackage{pifont}

\newcommand{\tg}{Telegram\xspace}
\newcommand{\impr}{IMProxy\xspace}

\newcommand{\im}{SIM\xspace}
\newcolumntype{Y}{>{\centering\arraybackslash}X}

\begin{document}

\title{
Practical Traffic Analysis Attacks on \\Secure Messaging Applications}



%
\author{\IEEEauthorblockN{Alireza Bahramali\IEEEauthorrefmark{1},
Amir Houmansadr\IEEEauthorrefmark{1},
Ramin Soltani\IEEEauthorrefmark{2},
Dennis Goeckel\IEEEauthorrefmark{3}, and
Don Towsley\IEEEauthorrefmark{1}}
\IEEEauthorblockA{University of Massachusetts Amherst\\
\IEEEauthorrefmark{1}\{abahramali, amir, towsley\}@cs.umass.edu}
\IEEEauthorblockA{\IEEEauthorrefmark{2}ramin.soltani@gmail.com}
\IEEEauthorblockA{\IEEEauthorrefmark{3}dgoeckel@engin.umass.edu}}



\IEEEoverridecommandlockouts
\makeatletter\def\@IEEEpubidpullup{6.5\baselineskip}\makeatother
\IEEEpubid{\parbox{\columnwidth}{
    Network and Distributed Systems Security (NDSS) Symposium 2020\\
    23-26 February 2020, San Diego, CA, USA\\
    ISBN 1-891562-61-4\\
    https://dx.doi.org/10.14722/ndss.2020.24347\\
    www.ndss-symposium.org
}
\hspace{\columnsep}\makebox[\columnwidth]{}}

\maketitle

\begin{abstract}
Instant Messaging (IM) applications like Telegram, Signal, and WhatsApp have become extremely  popular in recent years.
Unfortunately, such IM services have been  targets of continuous  governmental surveillance and censorship, as these services are home to public and private communication channels on socially and politically sensitive topics.
To protect their clients,
  popular IM services deploy state-of-the-art encryption mechanisms.
In this paper, we show that despite the use of advanced encryption, popular IM applications leak sensitive information about their clients to adversaries who merely monitor their encrypted IM traffic, with no need for leveraging any software vulnerabilities of IM applications.
Specifically, we devise traffic analysis attacks that enable an adversary to identify  administrators as well as  members of target IM channels (e.g., forums) with  high accuracies.
We believe that our study demonstrates  a significant, real-world threat to the users of such services
given the increasing attempts by oppressive governments at cracking down controversial IM channels.

We demonstrate the practicality of  our traffic analysis attacks through extensive experiments on real-world IM communications. We show that standard  countermeasure techniques such as adding cover traffic  can degrade the effectiveness of the attacks we introduce in this paper.
We hope that our study will encourage IM providers to integrate effective  traffic obfuscation countermeasures into their software.
In the meantime, we have designed and deployed an open-source, publicly available  countermeasure system, called  \impr, that can be used by IM clients  with no need for any support from IM providers. We have demonstrated the effectiveness of \impr through experiments.
\end{abstract}



%



\section{Introduction}
Instant Messaging (IM) applications like Telegram, Signal, and WhatsApp have become enormously  popular in recent years. Recent studies estimate that over 2 billion  people use mobile IM applications
across the world \cite{IM-users}. IM services enable users to form private and public social groups and exchange messages of various types, including text messages, images, videos, and audio files.
In particular, IM applications are used extensively to exchange politically and socially sensitive content.
As a result,  governments and corporations increasingly monitor the communications made through  popular IM services~\cite{Telegram-russia-keys,telegram-iran-servers,admin-identifying,admin-identifying2}.

A notable example of oppressed IM services is Telegram~\cite{Telegram} with over 200 million users globally~\cite{num-of-users}, where a large fraction of its users come from countries with strict media regulations like Iran and Russia.
In particular, Telegram  is so popular in Iran that it has been  estimated  to consume more than 60 percent of Iran's Internet bandwidth \cite{bandwidth-telegram}.
Consequently, Iranian officials have taken various measures to monitor and block Telegram: from requesting Telegram to host some of its servers inside Iran to enable surveillance~\cite{telegram-iran-servers}, to requesting Telegram to remove controversial  political and non-political channels~\cite{telegram-iran-servers}.
Eventually, Iran blocked Telegram entirely in April 2018 due to Telegram's non-compliance. Despite this,  statistics suggest only a small decrease in Telegram's  Iranian users who connect to it through various kinds of VPNs~\cite{iran-vpn}.
%
Telegram has also been blocked in  Russia as Telegram operators refrained from handing over their encryption keys to Russian officials for surveillance~\cite{Telegram-russia-keys}.
Finally, in the light of Telegram's crucial role in recent Hong Kong protests, there are unconfirmed reports~\cite{tg-HK,tg-HK2} that  mainland Chinese and Hong Kong authorities may have attempted to discover Hong Kong protesters by misusing a Telegram feature that enabled them to map phone numbers to Telegram IDs.

\paragraphb{A Fundamental Vulnerability:}
Popular IM applications like Telegram, WhatsApp, and Signal deploy encryption (either end-to-end or end-to-middle) to secure user communications. We refer to such services as \emph{secure IM (\im)} applications.
In this paper, we demonstrate that \emph{despite their use of advanced encryption, popular IM applications leak sensitive information about their clients' activities to surveillance parties}.
Specifically, we demonstrate that  surveillance parties are capable of identifying members as well as  administrators\footnote{An administrator of an IM channel is a member who is privileged to post messages to that channel.}   of target IM communications (e.g., politically sensitive IM channels) with very high accuracies, and by only using low-cost traffic analysis techniques.
Note that our attacks are \emph{not} due to security flaws or  buggy software implementations such as those discovered previously~\cite{imleak1,imleak2,imleak3,whatsapp-flaw-2019}; while important, such security flaws are scarce, and  are immediately fixed by IM providers once discovered.
Instead, our  attacks  enable surveillance by \emph{merely watching encrypted IM traffic} of IM users, and assuming that the underlying IM software is entirely secure.
The key enabler of our attacks is the fact that major IM operators \emph{do not} deploy any mechanisms to  obfuscate traffic characteristics (e.g., packet timing and sizes), due to the impact of obfuscation on the usability and performance of such services.
We therefore argue that \emph{our attacks demonstrate a fundamental vulnerability in major in-the-wild IM services}, and, as we will demonstrate, they work against all major IM services.

We believe that our attacks present \emph{significant real-world threats} to  the users of believed-to-be-secure IM services, specially given escalating attempts by oppressive regimes to crack down on such services, e.g., the recent attempts~\cite{admin-identifying,admin-identifying2,tg-HK,tg-HK2} to identify and seize the administrators and members of controversial IM communications.


\paragraphb{Our Contributions:} We design traffic analysis attack algorithms for \im communications;  the objective of our attack is to \emph{identify the admins and/or members} of target \im communications.
What enables our attack is that, \emph{widely-used \im services do not employ any  mechanisms to obfuscate statistical characteristics of their communications}.

 We start by establishing a statistical model for IM traffic characteristics.
Such a model is  essential in our search for effective traffic analysis attacks on \im services.
To model IM communications, we join over 1,000 public Telegram channels and record their communications, based on which we derive a statistical model for IM traffic features.
We use our \im model  to derive theoretical bounds on the effectiveness of our traffic analysis algorithms; we also use our statistical model to generate arbitrary numbers of synthetic \im channels to enhance the confidence of  our
 empirical evaluations.


Based on our statistical model for IM communications, we use hypothesis testing~\cite{poor2013introduction} to \emph{systematically} design effective traffic analysis attack algorithms. Specifically,
we design two traffic analysis attack algorithms;
our first algorithm, which we call the \emph{event-based} correlator, relies on the statistical model that we derive for \im communications to offer an optimal matching of users to channels.
Our second algorithm, which we call the \emph{shape-based} algorithm,  correlates the shapes of \im traffic flows  in order to match users to target channels. Our shape-based algorithm is slower but offers  more accurate detection performance than the event-based algorithm.
In practice, the adversary can  cascade the two algorithms  to optimize computation cost (and scalability) versus detection performance.
Note that, as demonstrated through experiments, our statistical detectors outperform
deep learning based detectors trained on IM traffic.
This is because, as also demonstrated in recent work~\cite{Nasr:2018:DSF:3243734.3243824}, deep learning traffic classifiers  outperform statistical classifiers \emph{only} in network applications with non-stationary noise conditions (e.g., Tor), where  statistical models becomes unreliable.

We perform extensive experiments on live and synthetic \im traffic to evaluate the performance of our attacks. We demonstrate that our algorithms offer extremely high accuracies in disclosing the participants of target \im communications. In particular, we show that \emph{only 15 minutes} of \tg traffic suffices for our shape-based detector to identify the admin of a target \im channel with a 94\% accuracy  and  a $10^{-3}$ false positive rate\textemdash the adversary can reduce the false positive to $5\times 10^{-5}$ by observing an hour of traffic (the adversary can do this hierarchically, e.g., by monitoring the users flagged using  15 min of traffic for longer traffic intervals).

We also study the use of standard traffic analysis countermeasures against our attacks. In particular, we investigate tunneling \im traffic through VPNs, mixing it with background traffic,  adding cover IM traffic, and delaying IM packets.
As expected, our experiments show that such countermeasures reduce the effectiveness of the attacks at the cost of additional communication overhead as well as increased latency for \im communications.
For instance, we find that tunneling \im traffic through VPN and mixing it with background web-browsing traffic  reduces the accuracy of our attack from 93\% to 70\%, and adding cover traffic  with a 17\% overhead  drops the accuracy to 62\%.
 We argue that since many  \im users do not deploy such third-party countermeasures due to usability reasons, \im providers should integrate standard traffic obfuscation techniques into their software to protect their users against the introduced traffic analysis attacks.
 In the meantime, \emph{we have designed and deployed an open-source, publicly available  countermeasure system, called  \impr, that can be used by IM clients  with no need to any support from IM providers}. We have demonstrated the effectiveness of \impr through experiments.

In summary, we make the following  contributions:
\begin{compactitem}
\item We introduce traffic analysis attacks that reliably identify  users involved in sensitive  communications through secure IM services. To launch our attacks, the adversary does not need cooperate with IM providers, nor does he need to leverage any security flaws of  the target IM services.
	\item We establish a statistical model for regular IM communications by analyzing  IM traffic from a  large number of real-world IM channels.
	\item We perform extensive experiments on the popular IM services of Telegram, WhatsApp, and Signal to demonstrate the in-the-wild  effectiveness of our attacks.
	\item We study potential countermeasures against our attacks. In particular, we design and deploy   \impr, which is an open-source, publicly available  countermeasure system. \impr works for all major IM services, with no need to any support from  IM providers.
\end{compactitem}

\section{Background: Secure Instant Messaging (\im) Applications}\label{sec:backg:im}

 We define  a \textbf{secure IM  (\im)} service to be an instant messaging service that satisfies two properties:
 (1) it deploys strong encryption on its user communications (either end-to-end or end-to-middle), and (2) it is  not controlled or operated by an adversary, e.g., a  government.
While our attacks also apply to non-secure IM  applications, an adversary can use other trivial techniques to compromise  privacy of  non-secure IM services.
For instance, if  the  operator of an IM service fully cooperates with a surveillant government, e.g., the \emph{WeChat} IM service in China or the \emph{Soroush} service in Iran, the IM provider can let the adversary identify target users with no need for traffic analysis mechanisms.   Similarly, an IM service with weak encryption can be trivially eavesdropped with no need for sophisticated traffic analysis attacks.

Table~\ref{sim-stats}  overviews some of the most popular \im services.

\begin{table*}
\centering
\caption{Popular IM services \cite{SIM-stats}}
\resizebox{0.8\linewidth}{!}{
\begin{tabular}{ |c|c|c|c|c|c| }
	\hline
	\textbf{IM Service} & \footnotesize \textbf{Monthly Users} & \textbf{Main Servers Hosted in} & \textbf{Owned by} & \textbf{End-to-End Encryption} & \textbf{Centralized} \\
	\hline
	WhatsApp & 1300 M & United States & Facebook & \checkmark & \checkmark \\
	\hline
	Facebook Messenger & 1300 M & United States & Facebook & \checkmark (Secret Communications) & \checkmark \\
	\hline
	WeChat & 980 M & China & Tencent & \xmark & \checkmark \\
	\hline
	QQ Mobile & 843 M & China & Tencent & \checkmark & \checkmark \\
	\hline
	Skype & 300 M & Estonia & Microsoft & \checkmark & \checkmark \\
	\hline
	Viber & 260 M & Luxembourg & Rakuten & \checkmark Since 2016 & \checkmark \\
	\hline
	Snapchat & 255 M & United States & Snap Inc & \xmark & \checkmark \\
	\hline
	LINE & 203 M & Japan & Line Corporation & \checkmark & \checkmark \\
	\hline
	Telegram & 200 M & UAE & Telegram Messenger LLP & \checkmark (Secret Chats) & \checkmark \\
	\hline
	Signal & 10 M & United States & Open Whisper Systems & \checkmark & \checkmark \\
	\hline
\end{tabular} }
\label{sim-stats}
\end{table*}



\subsection{How \im Services Operate}\label{sim-operate}\label{why-telegram}

\paragraphb{Architecture:}
All major IM services are \emph{centralized}, as shown in Table~\ref{sim-stats}. Therefore,  all  user communications in such services are exchanged through servers hosted by the IM provider companies, e.g., Telegram Messenger LLP
(note that some less popular services use a peer-to-peer architecture, e.g., FireChat~\cite{fire-chat}, Ring~\cite{ring}, and Briar~\cite{briar}).
Each IM service has a server for authentication  and key exchange. A database server stores  message contents and other user information  (possibly encrypted with client keys). Some IMs use Content Delivery Networks (CDNs) to run their databases to improve quality of service and resist attacks.
Existing IM services use various  messaging protocols for user communications, including   Signal~\cite{10.1007/978-3-319-45982-0_22}, Matrix~\cite{matrix}, MTProto~\cite{mtproto}, and Off-the-Record~\cite{Borisov:2004:OCW:1029179.1029200}.
Each of these protocols involves several stages including authentication, key exchange, message transmission, re-keying, and MAC key publishing~\cite{johansen2017comparing}.


Popular IM services intermediate all user communications by having  user traffic go through their servers.
Such a centralized architecture  allows IM providers  to offer  high quality of services and solves critical issues like reaching to  offline clients and clients behind NAT/firewalls.
However,  this presents different  privacy threats to the users,
 as IM servers are involved in all user communications. Some IM services deploy end-to-end encryption to alleviate this, as presented below.

\paragraphb{Security Features:}
IM services use standard authentication mechanisms like authorization keys and public key certificates to \emph{authenticate} IM servers and peers~\cite{black1999umac, black2000cbc}.
Also, they use standard techniques to ensure the \emph{integrity} of messages.
All major IM services encrypt user communications to protect \emph{confidentiality}~\cite{Harris:2012:CAE:2555215}.
Some IM providers additionally deploy \emph{end-to-end encryption} on user communications. This prevents IM operators from seeing the content of communications; however, they can still see communication metadata, e.g., who is talking to whom and when.
WhatsApp, Skype, Line, as well as Telegram and Facebook Messenger offer end-to-end encryption, while WeChat, Snapchat, and the BlackBerry Messenger do not.

Also, several major IM providers, including WhatsApp, Viber, Signal, and Facebook Messenger, provide \emph{perfect forward secrecy} by using short-term session keys for user communications~\cite{johansen2017comparing}.
Please refer to Johansen et al.~\cite{johansen2017comparing}
for further discussion of other IM security features.


%




\subsection{Prior Security Studies of IM Services}\label{sec:other-attack-im}


\paragraphb{Metadata leakage:}
Coull and Dyer~\cite{Coull:2014:TAE:2677046.2677048} are the first to apply traffic analysis  on messaging applications.
They demonstrate traffic analysis attacks that can infer various meta-data of a target Apple iMessage user, specifically, the  operating system version, type of the IM action, and, to some degree, the  language of conversations.
More recently, Park and Kim~\cite{10.1007/978-3-319-31875-2_21} perform traffic analysis on the Korean KakaoTalk IM service,  to identify users' online activities using basic classification algorithms.
Our work is differ from these works in that the design of our detectors rely on theoretical foundations and meticulous modeling of IM communications. Also, we believe that our attacks are able to reveal  IM meta-data that is more sensitive  than what was identified by prior works. We demonstrate the applicability of  our attacks on several IM services, and design and evaluate tailored countermeasures.


%
%

\paragraphb{Security vulnerabilities:} Johansen et al.~\cite{johansen2017comparing}  surveyed different implementations of \im protocols such as Signal, WhatsApp, and Threema, and  evaluated their security and usability; they conclude that none of the studied applications are infallible.
Unger et al.~\cite{Unger:2015:SSM:2867539.2867716} performed a comprehensive study of instant messaging protocols focused on  their security properties around trust establishment, conversation security, and transport privacy. Also, Aggarwal et al.~\cite{8443844}  study the implementation of encryption in  widely-used messaging applications.

Furthermore, there have been various identity enumeration attacks on messaging applications. In particular,  as some IM services use  SMS text message to activate new devices, an adversarial phone company can initiate and intercept such authorization codes to either identify users or access their accounts. Alternatively,
 unconfirmed reports~\cite{tg-HK} suggest that  mainland Chinese and Hong Kong authorities may have attempted to discover Hong Kong protesters by misusing a Telegram feature that allowed one to discover the Telegram IDs of phone contacts (therefore, mapping phone numbers to their Telegram IDs); Telegram has promised to fix this issue through an update that will allow users to cloak their phone numbers~\cite{tg-HK2}.

Alternatively, Schliep et al.~\cite{Schliep:2017:BSM:3139550.3139568} evaluate the security of the Signal protocol against  Signal servers. They identify vulnerabilities that allow the Signal server to
learn the contents of attachments, re-order and drop
messages, and add/drop participants from group conversations. Note that their study targets  an entirely different adversary than ours, i.e., their adversary is a compromised/malicious Signal server, whereas in our case the adversary is any third-party who is able to  wiretap encrypted IM traffic. Also, their attacks only work against Signal, whereas our attacks apply to all major IM services as they rely on fundamental communication behavior of IM services.

\paragraphb{Communication privacy:}
The centralized nature of popular \im services makes them susceptible to various privacy issues.
First,  all user communications, including group communications and one-on-one communications, are established with the help of  the servers run by the  \im providers; therefore,  \im providers have access to the metadata of all communications, i.e., who is talking to whom, and channel ownership and membership relationships.
Recent works suggest using various cryptographic techniques, such as private set intersection, to protect privacy against the central operators, e.g., for contact discovery~\cite{kales2019mobile}.
Second, even if an IM service provider is not malicious, its servers may be compromised by malicious adversaries~\cite{compromised-server} or subpoenaed by governments, therefore putting client communication metadata at risk.

In traditional \im services, user communications are encrypted end-to-middle, i.e., between clients and \im servers. In such services, the \im providers can see not only the users' communication metadata but also their communication contents.
Recently, major \im providers such as Signal and WhatsApp have started to support end-to-end encryption, therefore protecting communication content from  \im providers.
Poor/buggy implementations of some \im services have resulted in various security flaws and meta-data leakage threats despite their use of end-to-end encryption~\cite{imleak1, imleak2, imleak3, imleak4, whatsapp-flaw-2019}, e.g., through on/off notifications in \tg~\cite{imleak2} and the recent WhatsApp vulnerability giving remote access to the hackers~\cite{whatsapp-flaw-2019}.



\paragraphb{Censorship:} The centralized architecture of popular \im services makes their censorship trivial: censors can easily blacklist a handful of IP addresses or DNS records to block all communications to a target \im service.
A straightforward countermeasure to unblock censored \im services is to use standard circumvention systems like VPNs~\cite{vpn} and Tor~\cite{Dingledine:2004:TSO:1251375.1251396}.
Alternatively, major \im services allow the use of circumvention proxies to evade blocking, e.g., as built into the recent versions of the Telegram software after censorship attempts by Iranian and Russian authorities.

\section{Attack and Threat Model}\label{attack-model}


In this work, we demonstrate \textbf{a fundamental attack} on IM services: \emph{our attacks are applicable to all major IM services, and are not due to  buggy software implementations that can be fixed through software updates}, as overviewed in Section~\ref{sec:other-attack-im}.
%
Our attacks are performed by an adversary who merely performs \emph{traffic analysis}. In this setting, the attacker does not need to compromise or coerce the \im provider, nor does she need to block the target IM service entirely. Instead, the adversary  performs traffic analysis to identify the participants of target IM communications in order to either punish the identified IM participants or  \emph{selectively block} the  target communications.
In particular, the adversary can use traffic analysis to  identify the administrators of controversial political or social IM channels and force them to shut down their channels (as seen in recent incidents~\cite{admin-identifying,admin-identifying2}). Alternatively, the adversary can use our traffic analysis attacks to identify the members of controversial IM channels, and thereby selectively disrupt the access to the target channels.

\subsection{Introducing the Players}

The \textbf{adversary} is a surveillance organization, e.g., an intelligence agency run by a   government. The  \textbf{goal
of the adversary}  is to \emph{identify (the IP addresses of) the members or administrators (owners) of
target IM conversations}.

A \emph{target IM conversation} can be a public IM channel (e.g., a chat room) on  politically or socially sensitive topics,
or a private IM conversation between target users, e.g., dissidents and journalists.

For the adversary to be able to conduct the attack,
she needs to be \emph{intercepting} the (encrypted) network traffic of the \textbf{monitored IM users}, e.g., by wiretapping the ISPs of the monitored users.
Therefore, considering  the Great Firewall of China as the adversary,   she can only perform the attack on the  IM users residing inside China.

\subsection{Threat Model}

We assume that the hosting IM service is a secure IM (\im) service, as defined in Section~\ref{sec:backg:im}.
Therefore, the adversary does not leverage any security vulnerabilities of the target \im service in performing the attack. For instance, the \im system does not leak the IP addresses (or other sensitive meta-data) of its clients to the adversary. Also, we assume all traffic between IM clients and the IM servers to be \emph{encrypted} with strong encryption.
Finally,  the operators of the \im service
\emph{do not cooperate with the adversary} in identifying  target members.

\subsection{How the Attack Is Performed }\label{ground-truth}

\begin{figure}[!t]
	\centering
	\includegraphics[width = \linewidth]{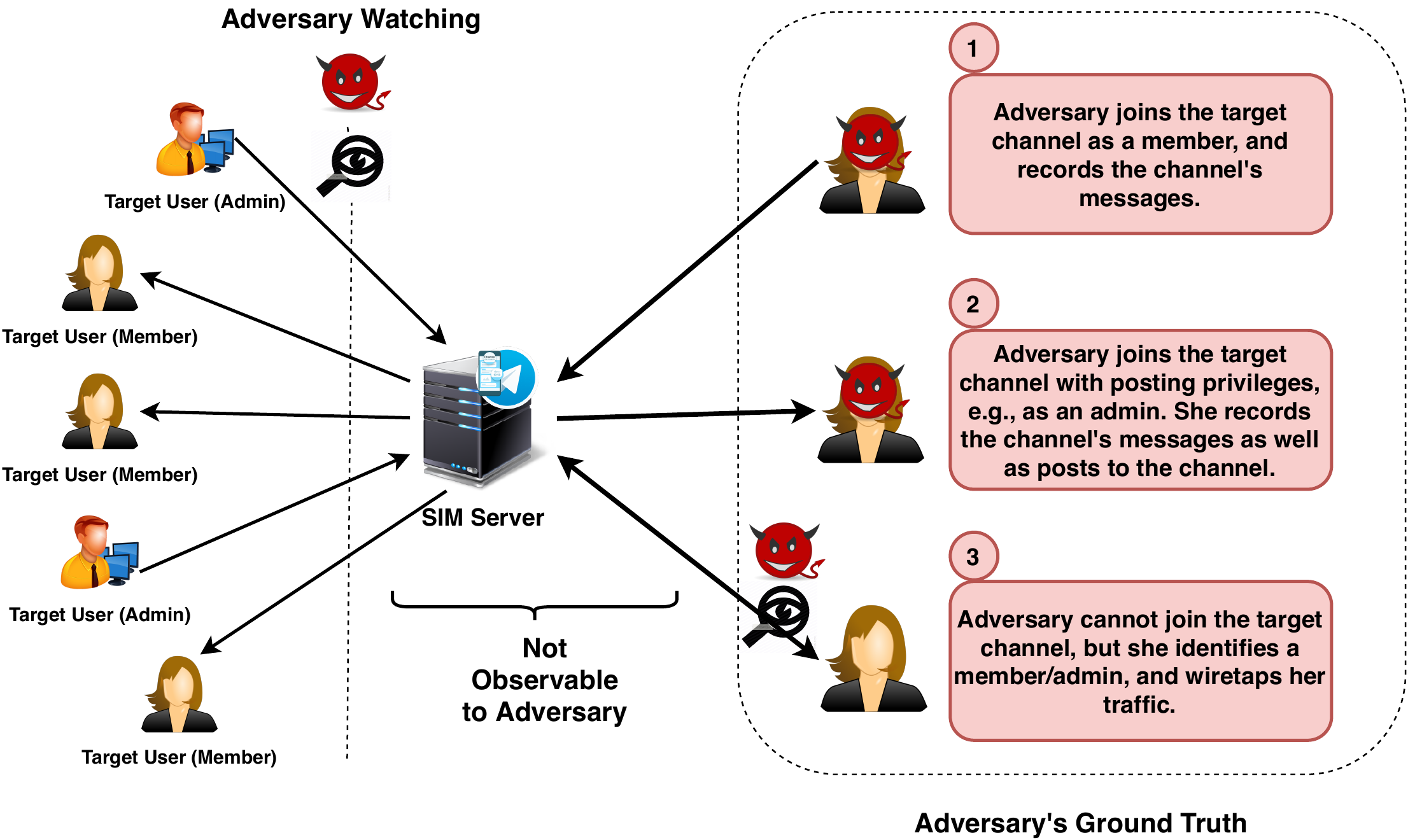}
	\caption{Alternative attack scenarios}\vspace{-3ex}
	\label{fig:attack-model}
\end{figure}

Figure~\ref{fig:attack-model} illustrates the setup of the attack.
Suppose that the adversary aims at identifying the members/admins of a specific IM channel, $C$.

\paragraphb{Adversary's ground truth:} For any target channel $C$, the attacker needs to obtain some \emph{ground truth} about the traffic of the channel. This can be done in three ways:
\begin{compactitem}
\item[(1)] If $C$ is an open (public) channel, the adversary joins the channel (as a member) and records the messages sent on $C$ along with their metadata (e.g., time and size of the messages).
\item [(2)] The adversary has joined $C$ and is capable of posting messages to  $C$. This can happen if $C$ a closed group that gives every member the capability to post messages, or this could be because the adversary has gained an admin role for $C$ (e.g., the surveillance adversary has created a channel on a politically sensitive topic to identify target journalists, or the adversary has arrested the owner of the sensitive channel and is misusing her account). In this setting, not only the adversary can record the messages posted to $C$, but also he can post his own messages to $C$ with his desired (distinct) traffic patterns.

\item [(3)] The adversary is not able to join $C$ as a member/admin, but he has identified (the IP address of) one of the members/admins of $C$. The adversary wiretaps the (encrypted) network traffic of the identified member and records the traffic patterns of the identified member.
\end{compactitem}

\paragraphb{Adversary's wiretap:} The adversary monitors the (encrypted) network traffic of IM users  to identify (the IP addresses of) the members/admins of the target IM channel $C$. This can be performed by the adversary wiretapping the network traffic of the ISPs or IXPs he is controlling, e.g., by the Great Firewall of China. Alternatively, the adversary can wiretap the network traffic of specific individuals (e.g., suspected activists), perhaps after obtaining a wiretapping warrant.

\paragraphb{Adversary makes decisions:} The adversary  uses a detection algorithm (as introduced in Section~\ref{detection-algo}) to match the traffic patterns of the wiretapped users to the ground truth traffic patterns of the target channel $C$.

\subsection{Related Traffic Analysis Attacks}\label{related-work}

Prior work has studied various kinds of traffic analysis attacks in different contexts.

\paragraphb{Flow correlation} In this setting, the adversary tries to link obfuscated network flows by correlating their traffic characteristics, i.e., packet timings and sizes~\cite{danezis2004traffic, donoho:raid02, he:tosp07, tor_cell, nasr2017compressive, shmatikov2006timing, wang:esorics02, zhang:sec00}. Flow correlation has particularly been studied as an attack on anonymity systems like Tor~\cite{back:ih2001, murdoch2005low, ramsbrock2008first, Wang2005, zhu:pet05}: the adversary can link the ingress and egress segments of a Tor connection (say, observed by malicious Tor guard and exit relays) by correlating the traffic characteristics of the ingress and egress segments. Recently, Nasr et al.~\cite{Nasr:2018:DSF:3243734.3243824} introduce a deep learning-based technique called DeepCorr which learns a correlation function to match Tor flows, and outperforms the previous statistical techniques in flow correlation.
Alternatively, flow correlation has been studied as a defensive mechanism to detect stepping stone attackers~\cite{he:tosp07,donoho:raid02}.

\paragraphb{Flow watermarking} This is the active version of flow correlation attacks described above. In flow watermarking, the adversary encodes an imperceptible signal into traffic patterns by applying slight perturbations to traffic features, e.g., by delaying packets~\cite{houmansadr2013need, houmansadr:ndss09, pyun:infocom07, wang:oakland07, yu:oakland07}.
Compared to regular (passive) flow correlation techniques, flow watermarks offer higher resistance to noise, but require real-time modification of network traffic, and are subject to detection attacks.

\paragraphb{Website fingerprinting} In Website Fingerprinting (WF), the adversary intercepts network connections of some monitored users and tries to match the patterns of the intercepted connections to a set of target webpages. This differs with flow correlation in that flow correlation intercepts the two ends of   target connections.
WF has particularly been studied as an attack on Tor. Existing WF techniques leverage various machine learning algorithms, such as k-NN, SVM, and deep neural networks to design classifiers that match monitored connections to target web pages~\cite{fing-attacks-defenses, k-fing, he2014novel, herrmann2009website, juarez2014critical, lu2010website, panchenko2016website, panchenko2011website, Wang:2014:EAP:2671225.2671235,DL-WF-18}.

\paragraphb{Intersection Attacks} Intersection attacks~\cite{kedogan2002limits, danezis2004statistical, donoho:raid02, 1199324} try to compromise anonymous communications by matching users' activity/inactivity time periods. For instance,  Kesdogan et al.~\cite{kedogan2002limits} model an anonymity system as an abstract threshold mix and  propose the disclosure attack whose goal is to learn the potential recipients for any target sender.  Danezis et al.~\cite{danezis2003statistical}  make the attack computationally more practical by proposing a statistical version of the attack.

\paragraphb{Side channel attacks} Another class of traffic analysis attacks aims at leaking sensitive information from encrypted network traffic of Internet services~\cite{skype,skype-revealing,chang2008inferring, Wright:2010:USP:1880022.1880029, Schuster:2017:BBR:3241189.3241295, gu2018walls, taylor2017robust,barradas2018effective}. For instance, Chang et al.~\cite{chang2008inferring}  infer speech activity from encrypted Skype traffic,  Chen et al.~\cite{chen2010side} demonstrate how online services leak sensitive client activities, and Schuster et al.~\cite{Schuster:2017:BBR:3241189.3241295} identify encrypted video streams.
%

\paragraphbe{Our Traffic Analysis Direction:} Our attacks presented in this paper are  \emph{closest in nature to the scenario of flow correlation techniques}. Similar to the flow correlation setting, in our scenario the adversary intercepts a live target flow (e.g., by joining a controversial IM channel), and tries to match it to the traffic patterns of flows monitored in other parts of the network  (to be able to identify the IP addresses of  the members or admins of the target channel). However, we can not trivially apply existing flow correlation techniques to the IM scenario, since  the traffic models and communication noise are entirely different in the IM scenario. We, therefore, design flow correlation algorithms tailored to the specific scenario of IM applications. To do so, we first model traffic and noise behavior in IM services, based on which we design tailored flow correlation algorithms for our specific scenario.

Note that one could alternatively use techniques from the intersection attacks literature to design traffic analysis attacks for IM services. However,  flow correlation is  significantly more powerful  than intersection attacks, as flow correlation  leverages not just the online/offline behavior of the  users, but also the patterns of their communications when they are online.
Also, typical IM clients tend to remain online for very long time intervals.
Therefore, we expect attacks based on  intersection to be significantly less reliable (or require very long observations to achieve comparable reliability) when compared to our  flow correlation-based attacks.

\section{Characterizing IM  Communications }\label{sec-model}

We start by characterizing IM traffic and deriving  a statistical model for it. We will use our model to find analytical bounds for the attack algorithms we design, as well as  to generate synthetic IM traffic to be used in some of our experiments.

\subsection{Main IM Messages}

IM services allow their users to send different types of messages, most commonly  text, image, video, file, and audio messages. IM messages are communicated between users through one of the following major communication forms:

\begin{compactitem}
	\item \textbf{Direct messages} are the one-on-one communications between IM users. As mentioned earlier, popular IM services are centralized, therefore all direct messages are relayed through the servers of the IM providers, and unless  end-to-end encryption is deployed, the servers can see communication contents.
	\item \textbf{Private (Closed) Group Communications}  are communications that happen between multiple users. In groups, every member can post messages and  read the messages posted by others. Each group has an administrator member who created the group and has the ability to manage the users and messages. An invitation is needed for a user to join a closed group.
	\item \textbf{Public (Open) Group Communications} which are also called \emph{channels}, are a broadcast form of communication in which one or multiple administrators can post messages, and the members can only read these posts.  Users can join channels with no need for an invitation.
\end{compactitem}

Note that  some IM services offer other forms of communications, like  status messages, that are not relevant to the attacks discussed in our work.

\subsection{Data Collection}\label{data}

In this paper, we collect the bulk of our IM data on the  \tg messaging application. We choose \tg for two reasons; first, \tg hosts a very large number of  \emph{public} channels that we can join to collect actual IM traffic. This is unlike other popular IM services where most group communications are closed/private. The second reason for choosing \tg for data collection is that \tg has been at the center of recent censorship and governmental surveillance attempts~\cite{admin-identifying, admin-identifying2, telegram-russia-servers, telegram-iran-servers}, as it is home to a multitude of politically and socially sensitive channels.
Note that our analysis and attacks are by no means specific to Telegram, as we demonstrate for other messaging services.

%

We use \tg's API to collect the  communications of 1,000  random channels with different message rates, each for a  24-hours span.
For every collected Telegram message, we extract the channel ID it was sent over, its timestamp, the type of message (text, photo, video, audio or file), and the message size.
\tg has a limit of 50 on the number of new channels a user can join every day. Therefore, we use multiple \tg accounts over several days to perform our data collection (also note that each \tg account needs to be tied to an actual mobile phone number, limiting the number of accounts one can create).

\begin{figure*}[!t]
\begin{center}
\begin{subfigure}[t]{0.23\textwidth}
	\centering
    \includegraphics[width=\linewidth, trim = {1cm 0cm 1cm 1cm}]{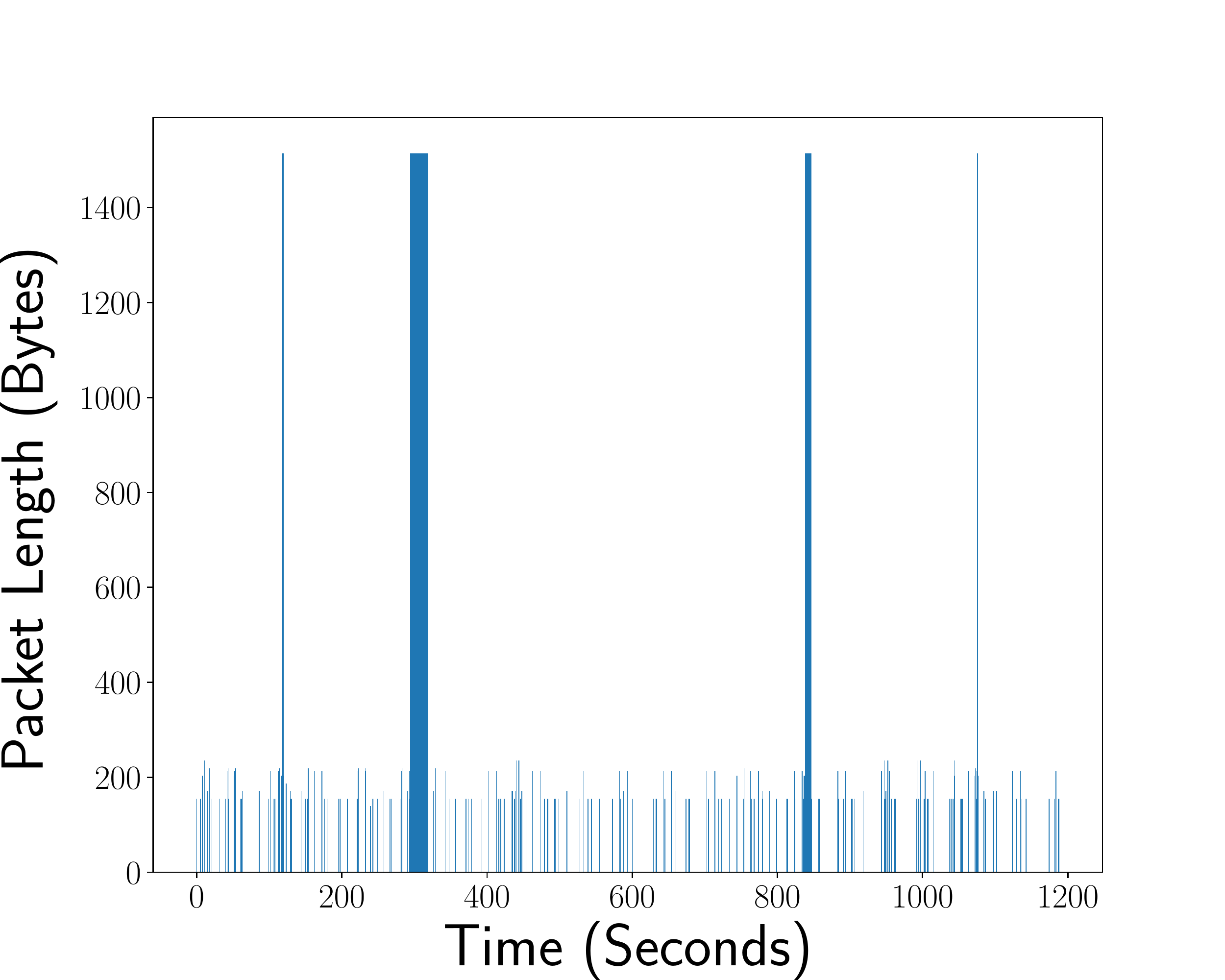}
    \caption{\tg}
\end{subfigure}
\begin{subfigure}[t]{0.23\textwidth}
	\centering
    \includegraphics[width=\linewidth, trim = {1cm 0cm 1cm 1cm}]{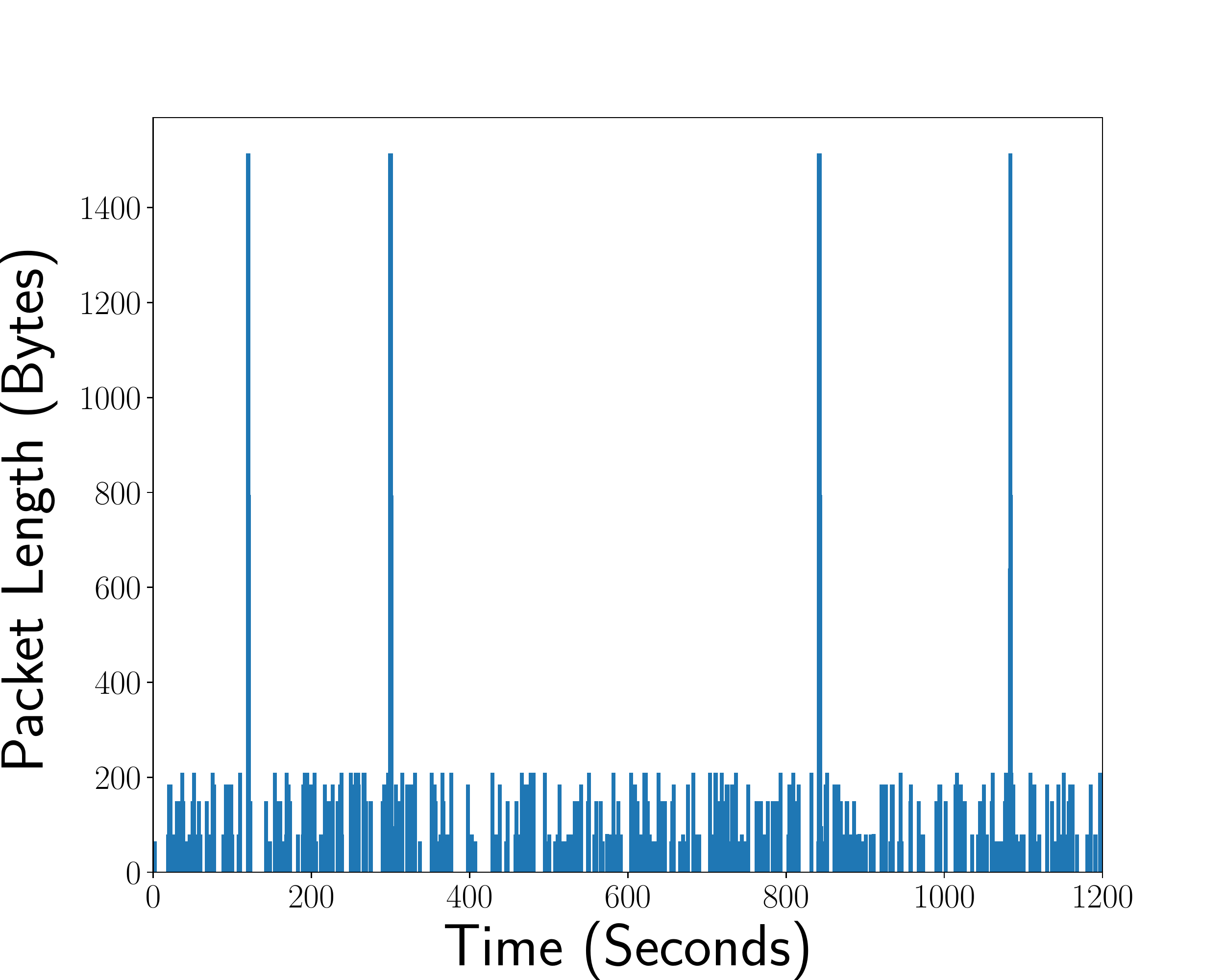}
    \caption{Signal}
\end{subfigure}
\begin{subfigure}[t]{0.23\textwidth}
	\centering
    \includegraphics[width=\linewidth, trim = {1cm 0cm 1cm 1cm}]{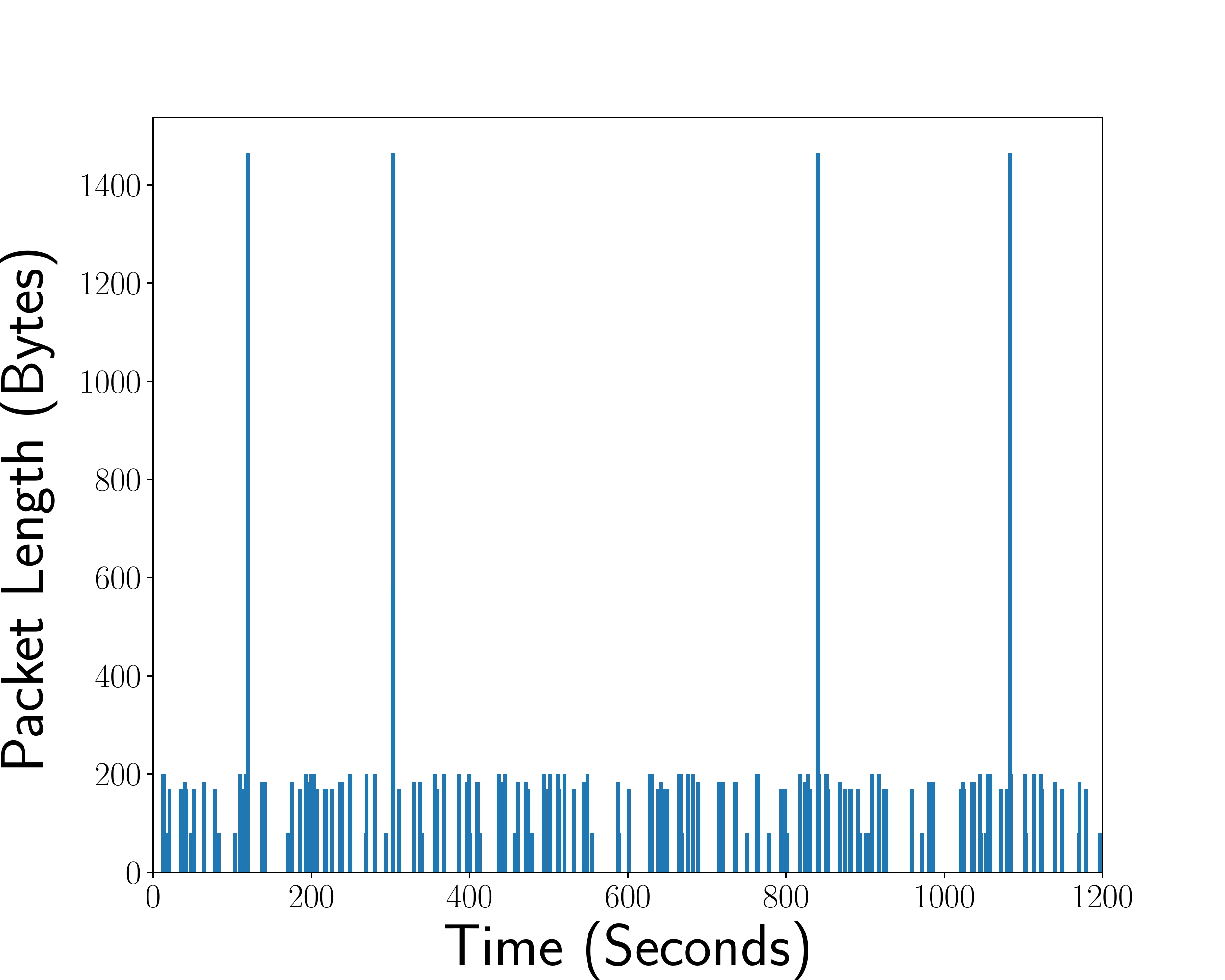}
    \caption{WhatsApp}
\end{subfigure}
\begin{subfigure}[t]{0.23\textwidth}
	\centering
    \includegraphics[width=\linewidth, trim = {1cm 0cm 1cm 1cm}]{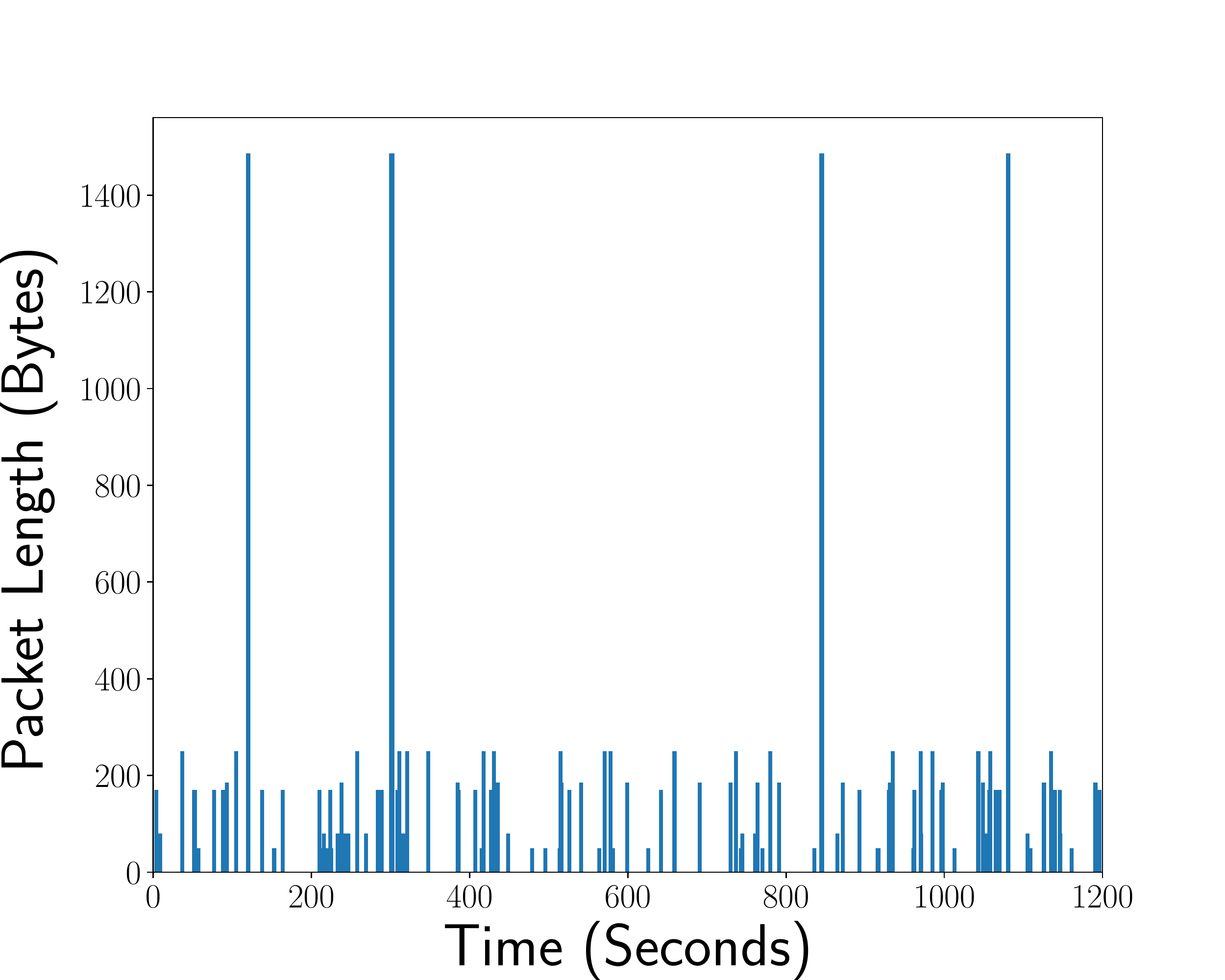}
    \caption{Viber}
\end{subfigure}
\end{center}
\vspace{-4ex}
\caption{Comparing the shape of traffic on four major \im services; by sending the same sequence of IM messages, we  observe similar traffic bursts regardless of the service provider. }
\label{other-sims}
\end{figure*}

Although we choose \tg for collecting \im traffic, we note that our attack algorithms perform similarly on  other \im{s} like WhatsApp and Signal. This is because none of these services implement traffic obfuscation, and therefore the shape of their traffic is similar across different IMs.
We have illustrated this in Figure~\ref{other-sims}, where  the same stream of messages are sent over four different \im services. As can be seen, the same messages result in similar traffic patterns across different IM services.


\subsection{Modeling IM Communications}\label{find-model}
We derive statistical models for IM communications based on our collected IM traffic. We model two key features of IM traffic: inter-message delays (IMDs) and message sizes. We also model the communication latency of IM traffic.
We use \textit{Maximum Likelihood Estimation (MLE)}~\cite{Pan2002} to fit the best  probability distribution  for each of these features.

\begin{figure*}[!htb]
\minipage{0.3\textwidth}
  \includegraphics[width=\linewidth, trim = {2cm 0cm 2cm 2cm}]{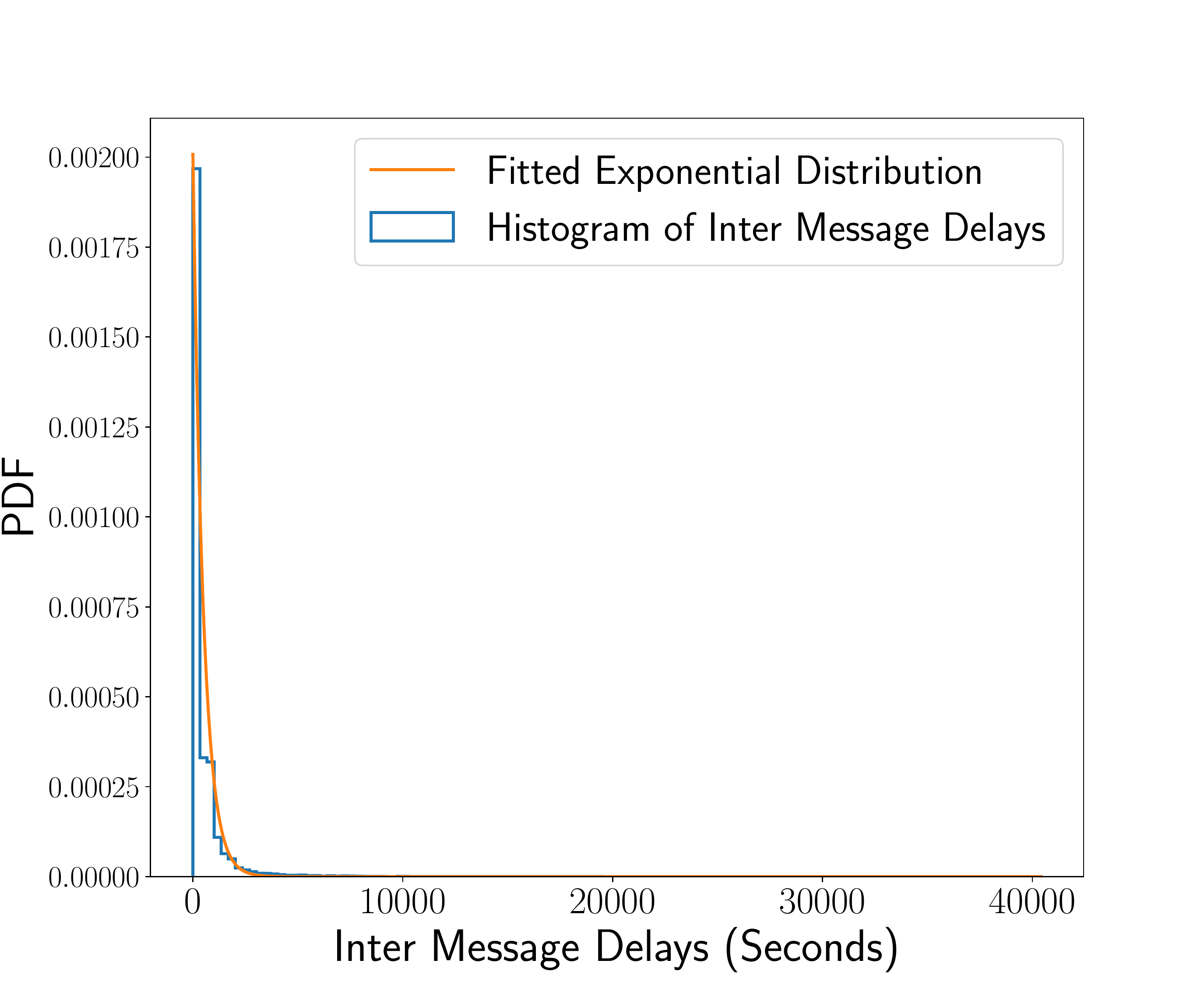}
  \caption{The PDF of inter-message delays and its fitted exponential distribution}
  \label{fig:time-dist}
\endminipage\hfill
\minipage{0.3\textwidth}
  \includegraphics[width=\linewidth, trim = {2cm 0cm 2cm 2cm}]{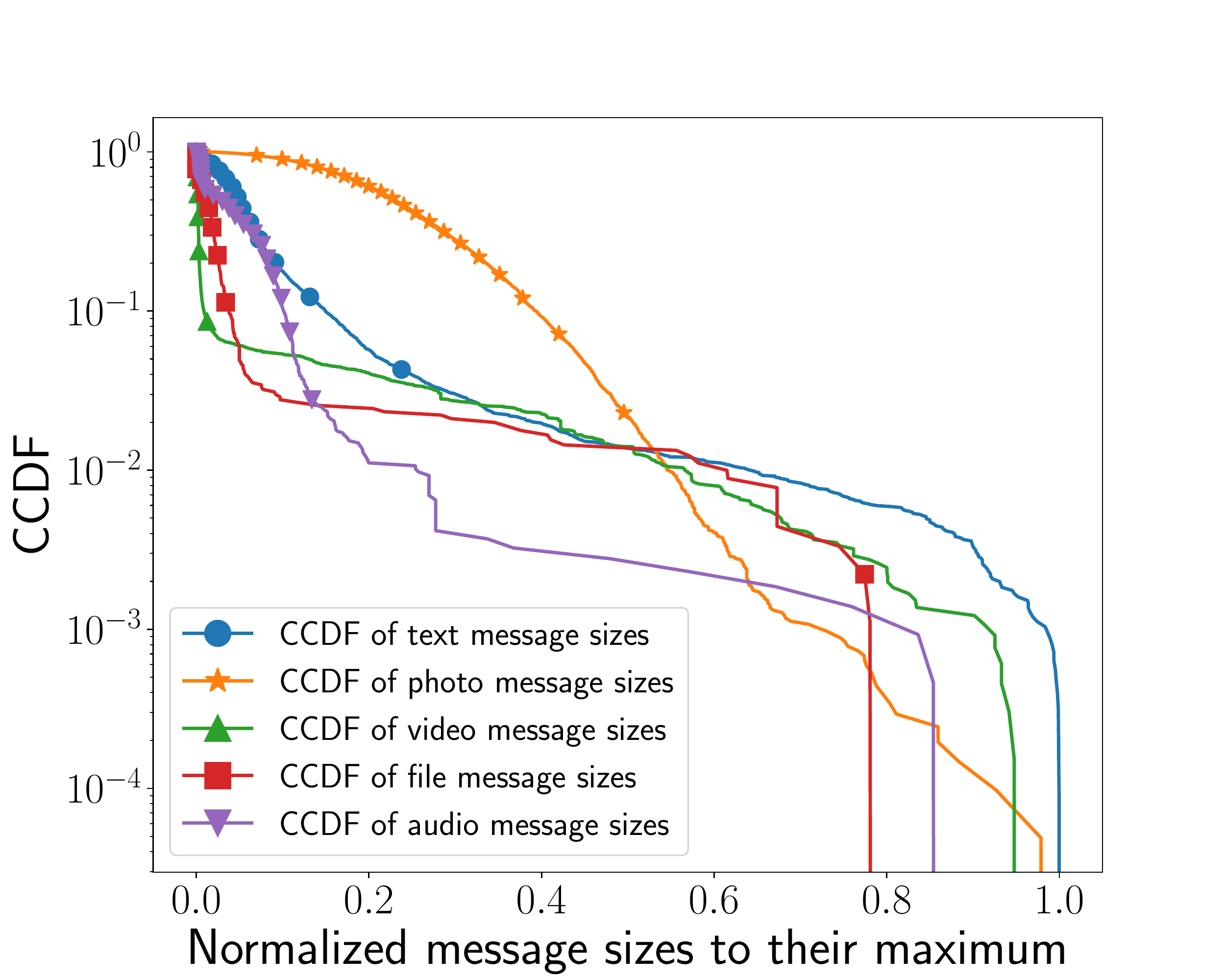}
  \caption{Complementary CDF (CCDF) of IM Size distributions for different types of messages}
  \label{fig:size-dist}
\endminipage\hfill
\minipage{0.3\textwidth}%
  \includegraphics[width=\linewidth, trim = {2cm 0cm 2cm 2cm}]{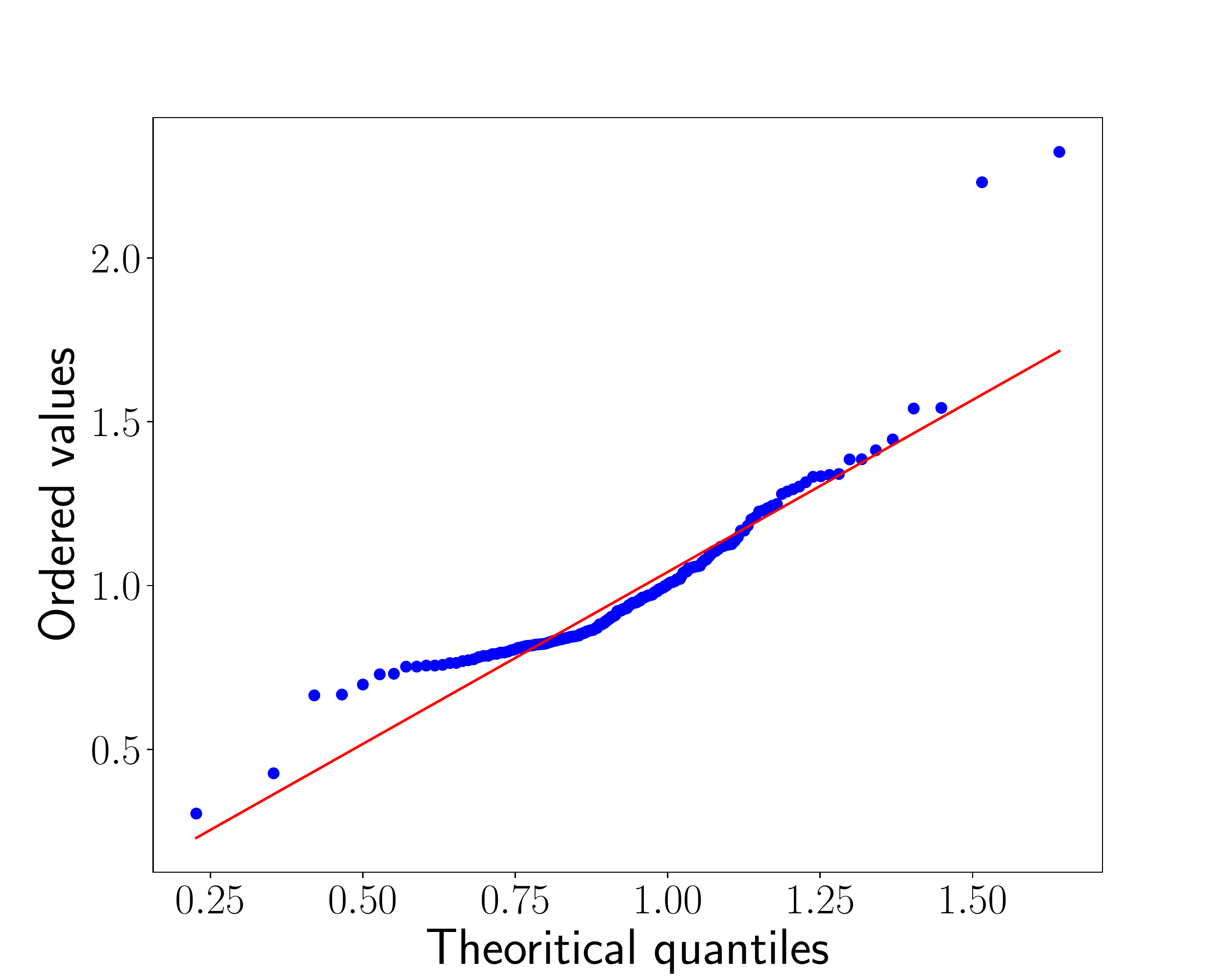}
  \caption{The Quantile-Quantile plot of transition delay and its fitted Laplacian distribution}
  \label{fig:qq-delay}
\endminipage
\end{figure*}

\paragraphb{Inter-Message Delays (IMDs):} The IMD feature is the time delay between consecutive IM messages in an IM communication. In our model, we merge  messages sent with very small IMDs (specifically,  messages separated by less than a threshold, $t_{e}$ seconds). We do this  because such extremely close messages create a combined traffic burst in the encrypted IM traffic that cannot be separated by the traffic analysis adversary.  Such close messages appear (infrequently) when an administrator forwards a batch of IM messages from another group.
We also filter out the very long IMDs that correspond to long late-night inactivity  periods.


We show that the probability density function of IMDs can be closely  fitted to an exponential distribution using our MLE algorithm;
Figure~\ref{fig:time-dist} shows the probability density function of IMDs for 200 IM channels with a message rate of 130 messages per day.  We interpret the exponential behavior of the IMDs to be due to the fact that messages (or message batches) are sent independently in the channels (note that this will be different for interactive one-on-one chats).

Also, we consider IMDs to be independent of the type and  size of messages, since in practice there is no correlation between the time  a message is sent and its type or size.




{\tiny
\begin{table}[t!]
\caption{Distribution of various message types}
\centering
\resizebox{\linewidth}{!}{
\begin{tabular}{ |c|c|c|c|c| }
	\hline
	\textbf{Type} & \textbf{Count} & \textbf{Volume (MB)} & \textbf{Size range} & \textbf{Avg. size}\\
	\hline
	Text & 12539 (29.4\%) & 3.85 (0.016\%) & 1B-4095B & 306.61B\\
	\hline
	Photo & 20471 (48\%) & 1869.57 (0.765\%) & 2.40Kb-378.68Kb & 91.33KB\\
	\hline
	Video & 6564 (15.4\%) & 232955.19 (95.3\%) & 10.16Kb-1.56Gb & 35.49MB \\
	\hline
	File & 903 (2.1\%) & 47.46 (0.019\%) & 2.54Kb-1.88Mg & 52.56KB \\
	\hline
	Audio & 2161 (5.1\%) & 9587.36 (3.92\%) &  2.83Kb-98.07Mg & 4.44MB \\
	\hline
\end{tabular} }
\label{messages-table}
\end{table}
     }

\begin{figure}[!t]
     \centering
     \includegraphics[width = 0.9\linewidth]{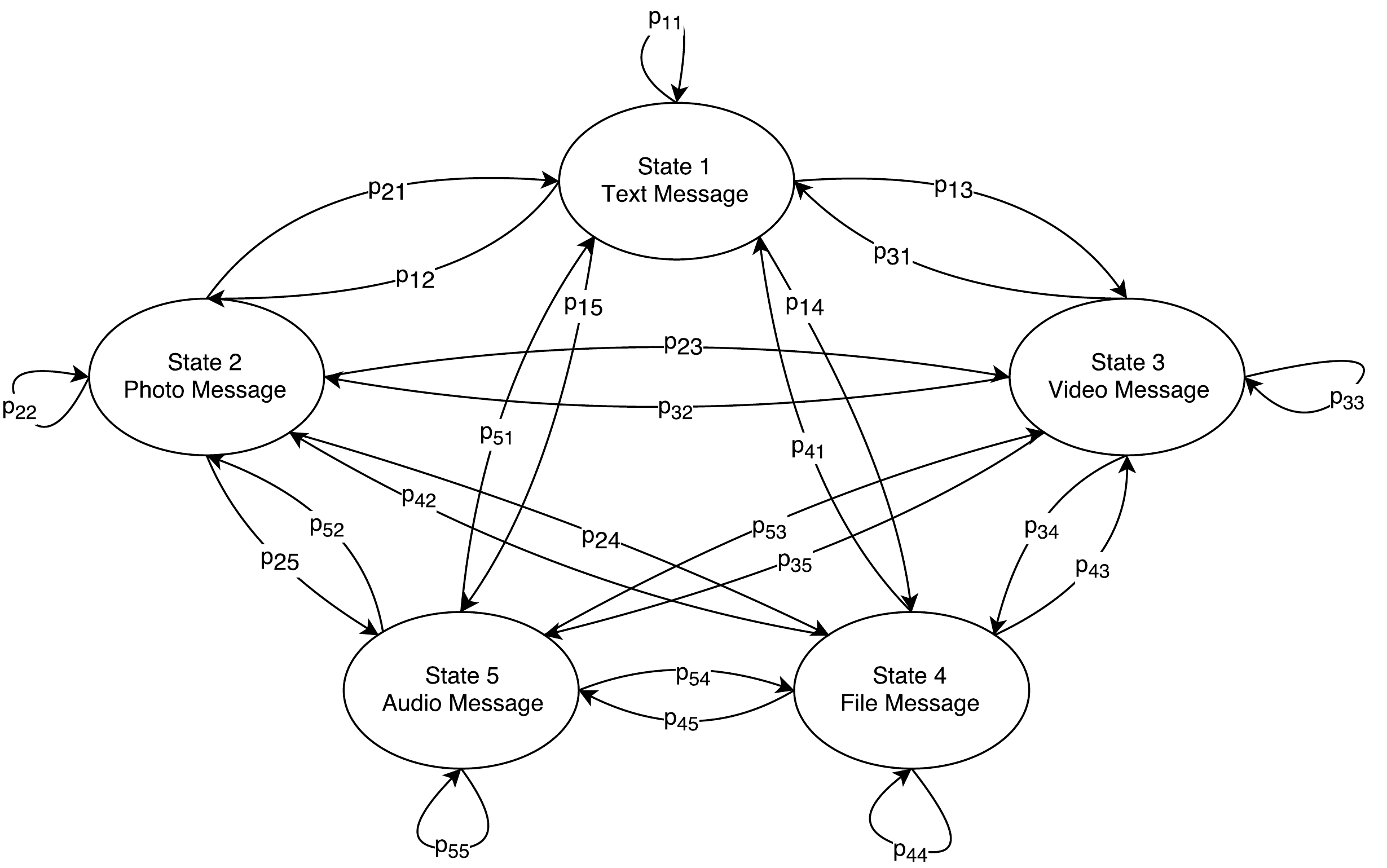}
     \caption{Markov chain of IM message sizes}\vspace{-4ex}
     \label{fig:markov}
\end{figure}

\paragraphb{Messages Sizes:} Table~\ref{messages-table} shows the size statistics and frequencies of the five main message types in our collected IM messages.
We use these empirical statistics to create a five-state Markov chain, shown in  Figure~\ref{fig:markov},  to model the sizes of the messages sent in an IM communication stream.
We obtain the empirical transition probability matrix of this Markov model for the aggregation of all channels, as well as for groups of channels with similar rates; the matrices are presented in Appendix~\ref{tr}.
We see that the transition matrices change slightly for IM channels with different daily message rates.

Finally, Figure~\ref{fig:size-dist} shows the Complementary Cumulative Density Function (CCDF) of the normalized message sizes for different message types (the sizes are normalized  by the maximum message size of each category). We observe that  different message types are characterized by  different message size distributions.

\paragraphb{Communication Latency:}
IM messages are delayed in transit due to two reasons: network latency and the IM servers' processing latency.
To measure such latencies, we collect IM traffic from 500 channels, each for one hour (therefore, 500 hours worth of IM traffic) using \tg's API. We then set up two IM clients, and send the collected IM traffic between the two clients to measure the incurred communication latencies.
%
Using MLE, we find that transition latencies   fit best to  a Laplacian distribution $f_{\mu,b}(x)$, where $\mu$ is the average and $2b^{2}$ is the variance of the delay.
Since network delay  cannot be negative, we consider only the positive parts of the Laplace distribution. Figure~\ref{fig:qq-delay} shows a Quantile-Quantile (Q-Q) plot of the packet latencies against the best Laplace distribution.

\subsection{Synthesizing \im Traffic}\label{synthesize}
We can use our empirical model, described above, to create synthetic IM communications. A synthetic IM communication trace consists of IM messages with specific sizes, timings, and message types. Such synthetic traffic allows us to validate our experimental results using much more traffic samples than what can be collected online.

The sketch of our algorithm in presented in Algorithm~\ref{syn-alg}.
The inputs of the algorithm are $\lambda$, the rate of messages (per day) for the channel to be synthesized, and $T$, the length of  the synthesized channel.
First, the algorithm uses the empirical distribution of IMDs (shown in Figure~\ref{fig:time-dist}) to create a sequence of IM message timings. Then,
our algorithm uses our Markov model for message types to assign a type to each of the messages in the sequence (Section~\ref{find-model}).
Finally, for each message, our algorithm finds its size using the empirical distribution of sizes for the corresponding type of the message (i.e., from Figure~\ref{fig:size-dist}).
The output of the algorithm is a sequence of IM messages, where each message has a timestamp, a  size, and a message type.

\begin{algorithm}[!t]
\caption{Algorithm for Generating Synthetic IM  Traffic}\label{syn-alg}
\begin{algorithmic}[1]
\Procedure{GenerateSyntheticTraffic}{}
\State $\textit{P} \gets \text{transition matrix based on}$ $\lambda$
\State $\textit{current length} \gets 0 $
\State $\textit{message sequence} \gets \varnothing $
\While{$\textit{current length} < \textit{T}$}
\State $\textit{event} \gets \varnothing $
\State $\textit{t} \gets \text{A random time from IMD}$
\State \hspace{\algorithmicindent} empirical distribution
\State $\textit{current length} \gets \textit{current length} + \textit{t}$
\State $\textit{event} \gets \textit{event} + \{\textit{t}\}$
\State $\textit{message sequence} \gets \textit{message sequence} + \{\textit{event}\}$
\EndWhile
\For{\text{each} \textit{event} \text{in} \textit{message sequence}}
\State $\textit{event} \gets \textit{event} + \text{\{\textit{message type}\} based on \textit{P}}$
\State $\textit{size} \gets \text{A random size from the corresponding}$
\State \hspace{\algorithmicindent} message type empirical distribution
\State $\textit{event} \gets \textit{event}+\{\textit{size}\}$
\EndFor
\State \Return {$\textit{message sequence}$}
\EndProcedure
\end{algorithmic}
\end{algorithm}


Later in  Section~\ref{simulation} we show that our detectors provide comparable performances on synthetic and real-world IM traces (for similar settings), demonstrating the realisticity  of our synthetic traces.
Since  our  traffic synthesizing algorithm uses sample IM traces to generate synthetic IM traffic patterns, the quality   of its synthesized traffic improves by increasing the size of its training dataset.
Alternatively, one could train a generative adversarial network to produce synthetic IM traces; we leave this to future work.

\section{Details of Attack Algorithms}\label{detection-algo}

We design two algorithms for performing our attack (i.e., to map monitored IM users to their channels).
As discussed in Section~\ref{related-work}, our attack scenario is closest in nature to the scenario of flow correlation attacks. Therefore, the design of our attacks is inspired by existing work on flow correlation.
Prior flow correlation techniques use standard statistical metrics, such as mutual information~\cite{chothia2011statistical,zhu2004flow}, Pearson correlation~\cite{levine2004timing,shmatikov2006timing}, Cosine Similarity~\cite{nasr2017compressive,houmansadr2014non}, and the Spearman Correlation~\cite{raptor},  to link  network flows by correlating their vectors of packet timing and sizes.
We use \emph{hypothesis testing}~\cite{poor2013introduction},\footnote{Our approach is ``threshold testing'' by some of the more strict definitions,  however, we will use the term ``hypothesis testing'' in this paper as threshold testing  falls into the general class of statistical hypothesis tests~\cite{poor2013introduction}.}
similar to state-of-the-art flow correlation works~\cite{houmansadr2014non, houmansadr:ndss09},  to design optimal traffic analysis algorithms for   the particular setting of IM communications.
In contrast to flow correlation studies which use the  features of network packets, we use the features (timing and sizes) of \textit{IM messages} for detection.

Note that,
the recent work of  DeepCorr~\cite{Nasr:2018:DSF:3243734.3243824} uses a  deep learning classifier to perform flow correlation attacks on Tor. They demonstrate that their deep learning classifier outperforms statistical correlation techniques in linking Tor connections.
In Section~\ref{deepcorr}, we compare our statistical classifiers with a DeepCorr-based classifier tailored to IM traffic. As we will show,  our statistical classifiers \emph{outperform} such deep learning based classifiers, especially for shorter flow observations. Intuitively, this is due to the sparsity of  events in typical  IM communications, as well as the  stationary nature of noise in IM communications in contrast to the scenario of Tor. Note that this fully complies with Nasr et al.~\cite{Nasr:2018:DSF:3243734.3243824}'s observation that DeepCorr only  outperforms statistical  classifiers in non-stationary noisy conditions, where statistical traffic models become inaccurate.

\paragraphb{Our hypothesis testing:} Consider $C$ to be a target \im channel (e.g., a public group on a politically sensitive topic).
For each IM user, $U$, the attacker aims at deciding which of the following hypotheses is true:

\begin{compactitem}
    \item $H_{0}$: User $U$ is \emph{not} associated with the target channel $C$, i.e., she is neither a member nor an administrator of channel $C$.
    \item $H_{1}$: User $U$ \emph{is} associated with the target channel $C$, i.e., she is posting messages to that channel as an admin, or is a member of that channel and therefore receives  the channel's messages.
\end{compactitem}



As described in our threat model (Section~\ref{attack-model}), the adversary can only observe encrypted \im communications between users and \im servers. Therefore, we design detectors that use traffic features, i.e., IMDs and message sizes. In the following, we describe two detector algorithms.


\subsection{Event-Based Detector}\label{event-based}
\begin{figure}[]
     \centering
     \includegraphics[width = \linewidth, trim = {0cm 0cm 0cm 2.2cm}]{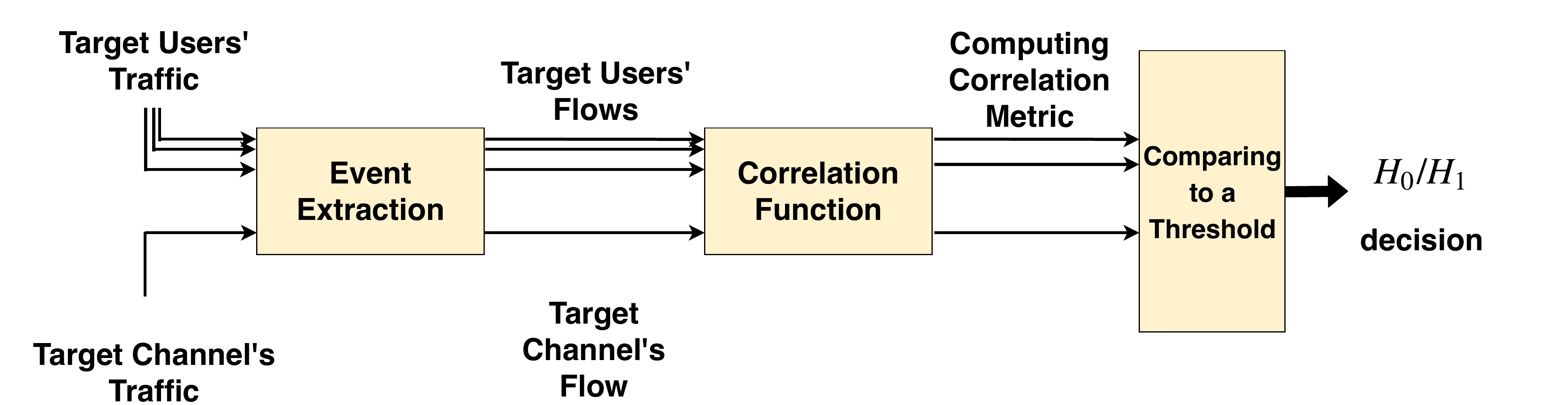}
     \caption{Event-based detector}\vspace{-3ex}
     \label{fig:size-model}
\end{figure}

\begin{figure}[!htbp]
\centering
\begin{subfigure}{\textwidth}
     \includegraphics[width = 0.5\linewidth]{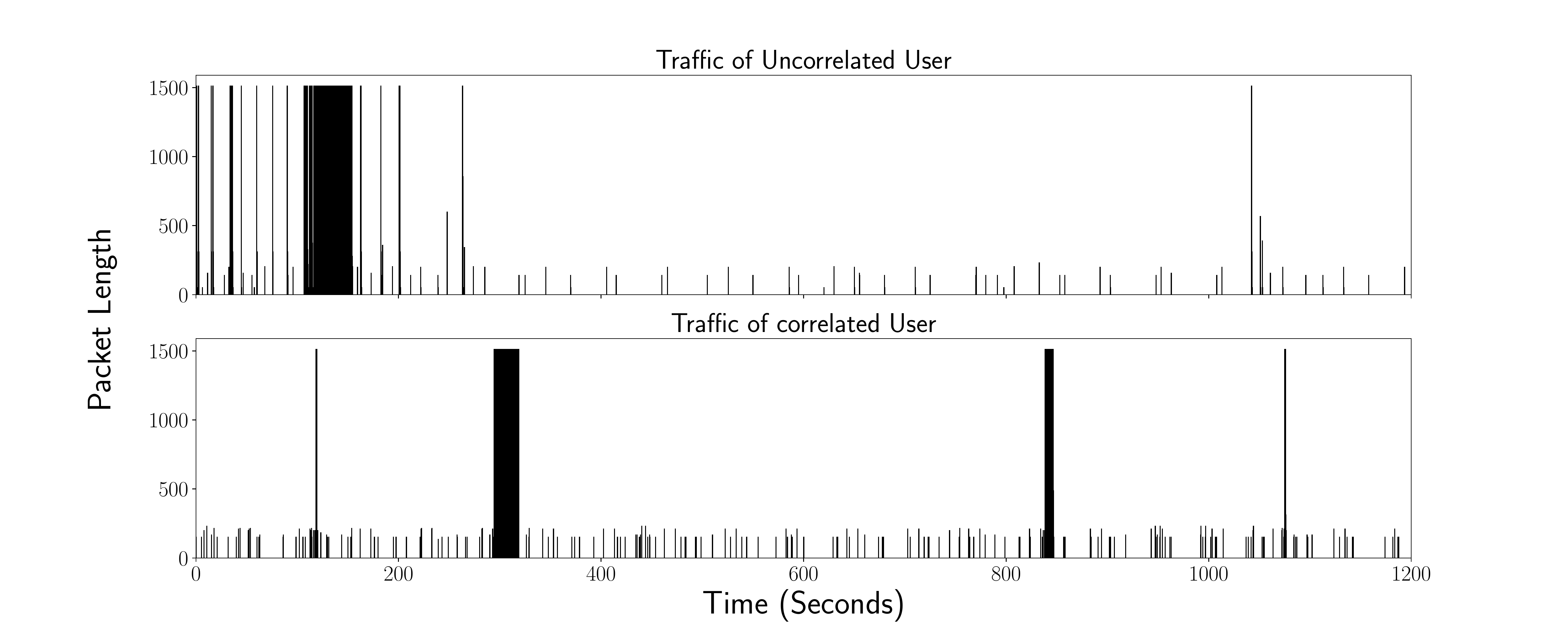}
\end{subfigure}
\begin{subfigure}{\textwidth}
     \includegraphics[width = 0.5\linewidth]{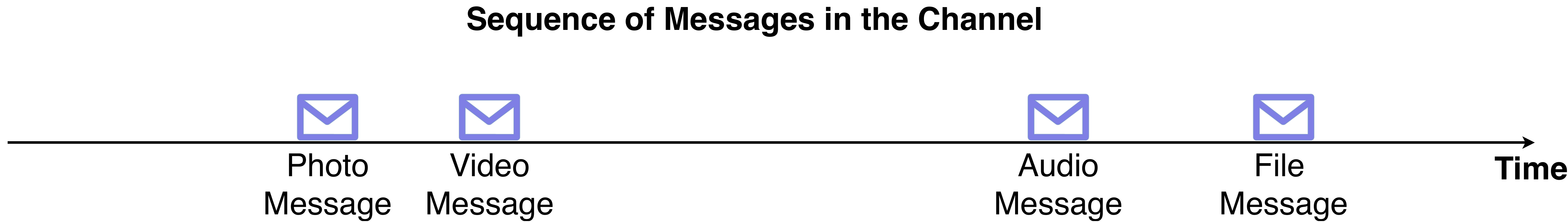}
\end{subfigure}
\caption{\small{Event extraction: IM Messages sent/received by a target user create bursts of (encrypted) packets; the adversary can extract events from packet bursts.} }\vspace{-3ex}
\label{fig:burst-user}
\end{figure}

Our first detector, the \emph{Event-Based Detector}, aims at matching \im events in a target user's traffic to those of the  target channel $C$.
An \emph{event} $e=(t,s)$ is a single \im message or a batch of \im messages sent with IMDs less than a threshold $t_{e}$ (as introduced earlier). Each single \im message can be one of the five types of image, video, file, text, or audio. $t$ is the time that $e$ appeared on the \im communication (e.g., sent to the public channel), and $s$ is the size of $e$.
Note that an \im communication can include \im protocol messages as well (handshakes, notifications, updates, etc.); however, such messages are comparatively very small as shown in Figure~\ref{fig:burst-user}, and thus the detector ignores them in the correlation process.
Recall that the adversary is not able to see plaintext events in the user's traffic due to encryption. Therefore, the first stage of our event-based detector is to \emph{extract} events based on the user's encrypted \im traffic shape.
Figure \ref{fig:size-model} depicts the components of our event-based detector.


\paragraphb{Event Extraction:}
%
Each \im event, e.g., a sent image, produces a burst of MTU-sized packets in the encrypted traffic, i.e., packets with very small inter-packet delays.
This is illustrated in Figure~\ref{fig:burst-user}: \im events such as images appear as traffic bursts, and scattered packets of small size are \im protocol messages like notifications, handshakes, updates, etc.
Therefore, the adversary can extract \im events by looking for bursts of MTU-sized  packets, even though she cannot see packet contents due to encryption.
We use the IPD threshold $t_{e}$ to identify bursts. Any two packets with distance less than $t_{e}$ are considered to be part of the same burst.
Note that $t_{e}$ is a hyper-parameter of our model and we  discuss its choice later in the paper.
For each burst, the adversary extracts a \im event, where the arrival time of the last packet in the burst gives the arrival time of the event, and the sum of all packet sizes in the burst gives the size of the event.
Two \im messages sent with an IMD less than $t_{e}$ are extracted as one event. Similarly, the adversary combines events closer than $t_{e}$ when capturing them from the target channel.



\paragraphb{Forming Hypotheses:}
We call a one-sided \im communication a \im \emph{flow}. Therefore, a flow either consists of the packets \emph{sent} by a  user  to an \im server, or the packets \emph{received} by the user from the \im server.
We represent  a flow with $n$ events as  $f = \{e_{1}, e_{2},\ldots,e_{n}\}$, where $e_i = (t_i,s_i)$ is the $i$th event.

Consider a user $U$ and a target channel $C$.
Suppose that the adversary has extracted flow $f^{(U)} = \{e^{(U)}_{1}, e^{(U)}_{2},\ldots,e^{(U)}_{n}\}$ for user $U$ (through wiretapping), and flow $f^{(C)} = \{e^{(C)}_{1}, e^{(C)}_{2},\ldots,e^{(C)}_{n}\}$ for the target channel $C$ (using her ground truth). The detector aims at deciding  whether user $U$ is an administrator or member of the channel.
We can re-state the adversary's hypotheses  presented earlier in this section as follows:
\begin{compactitem}
\item $H_{0}$: User $U$ is not an administrator or member of the target channel; hence, $f^{(C)}$ and $f^{(U)}$ are independent.
\item $H_{1}$: User $U$ is an administrator or member of the target channel $C$; therefore, the user flow $f^{(U)}$ is a noisy version of the channel flow $f^{(C)}$.
\end{compactitem}

\noindent
Therefore, we have\footnote{Note that the above hypothesis is for the case that the adversary is looking for the ``administrator'' of channel $C$.
For the case that the adversary is looking for the ``members'' of the target channel, the hypothesis changes slightly by replacing  $t_i^{(C)}=t_i^{(U)}+d_i^{(U)}$ with $t_i^{(U)}=t_i^{(C)}+d_i^{(C)}$. As the derivations will be  exactly the same, we exclude it without loss of generality. }
\label{h-test}
\[	\begin{cases}
	H_{0}: t_i^{(C)}=t_i^{(*)}+d_i^{(*)},s_i^{(C)}=s_i^{(*)}, 1 \leq i \leq n  \\
	H_{1}: t_i^{(C)}=t_i^{(U)}+d_i^{(U)},s_i^{(C)}=s_i^{(U)},  1 \leq i \leq n
	\end{cases}\]
where $f^{(*)} = \{e^{(*)}_{1}, e^{(*)}_{2},\ldots,e^{(*)}_{n}\}$ is the flow of a user $U^\prime\neq U$ who is \emph{not} an  administrator/member of channel $C$.
Also, $d_{i}^{(\cdot)}$ is the latency applied to the timing of the $i$th event. Note that IM message sizes do not change drastically in transit, and the order of messages remains the same after transmission.


\paragraphb{Detection Algorithm:} The adversary counts the number of event matches between  the user flow $f^{(U)}$ and the channel flow $f^{(C)}$. We say that the $i$th channel event $e_i^{(C)}$ matches \emph{some} event $e_j^{(U)}$ in $f^{(U)}$ if:
\begin{compactitem}
	\item $e_i^{(C)}$ and $e_j^{(U)}$ have close timing: $|t_{i}^{(C)}- t_{j}^{(U)}|<\Delta$; and
	\item $e_i^{(C)}$ and $e_j^{(U)}$ have close sizes: $|s_i^{(C)}-s_j^{(U)}| <\Gamma$.
\end{compactitem}
where $\Delta$ and $\Gamma$ are thresholds discussed in Section~\ref{param}.
Note that even though the sizes of \im messages do not change in transmission, the event extraction algorithm introduced earlier may impose size modifications, as network jitter  is able to divide/merge event bursts (i.e., a burst can be divided into two bursts due to network jitter or two bursts can be combined due to the small bandwidth of the user).

Finally, the adversary calculates the ratio of the matched events as $r=k/n$, where $k$ is number of matched events and $n$ is the total number of events in the target channel. The detector decides the hypothesis by comparing to a threshold: $r=\frac{k}{n}\underset{H_{0}}{\overset{H_{1}}{\gtrless}}\eta$.
The detection threshold $\eta$ is discussed in Section~\ref{result:realworld}.

\paragraphb{Analytical Bounds:}
We first derive an  upper-bound on the probability of false positive ($\mathbb{P}_{\mathrm{FP}}$), i.e.,  the probability that $H_1$ is detected when $H_0$ is true (Type I error). Let $p_{0}$ be the probability that a message with size $s_i^{(C)}$ and time $t_i^{(C)}$ matches an event in $f^{(U)}$ when $H_0$ is true, i.e., there exists only one message whose time $t_j^{(*)}$ satisfies $t_i^{(C)}\leq t_j^{(*)} \leq t_i^{(C)} + \Delta$ and has the same size label as $s_i^{(C)}$. From our observations, $p_{0}=0.002$. This Type I error occurs if more than $\eta\cdot  n$ events in $f^{(C)}$ match $f^{(U)}$, when $H_0$ is true. This is equivalent to the case that less than $n-\eta\cdot n$ events in $f^{(C)}$ do not match $f^{(U)}$ when $H_0$ is true. Consequently,
\begin{align}
\nonumber \mathbb{P}_{\mathrm{FP}} = \mathbb{P}(k \geq \eta n\mid H_0)&= \mathbb{P}(n-k \leq n- \eta n\mid H_0) ,\\
\nonumber &=F(n-\eta n;n,1-p_{0}),\\
 &\leq \left(\frac {1-\eta}{p_{0}}\right)^{-n+n \eta \eta} \left(\frac{\eta}{1-p_{0}}\right)^{-n\eta},
\label{fp}
\end{align}
where $F(r;m,p)=\mathbb{P}(X\leq r)$ is the cumulative density function of a Binomial distribution with parameters $m,p$, and the last step follows from the following inequality which is tight when $p$ is close to zero~\cite{Arratia1989}:
\begin{align}
\label{eq:1} F(r;m,p)  \leq \left(\frac{r/p}{p}\right)^{-k}  \left(\frac{1-r/m}{1-p}\right)^{k-m}
\end{align}
%
%
%

Next, we upper-bound the probability of false negatives ($\mathbb{P}_{FN}$), i.e.,  the probability that $H_0$ is detected when $H_1$ is true, which occurs when less than $k$ messages of $f^{(C)}$ match $f^{(U)}$. Let $p_{1}$ be the probability of the case that an event of $f^{(C)}$ matches $f^{(U)}$ when $H_1$ is true (Type II error).

Even though we mentioned earlier in this section that when $H_{1}$ is true, a delayed version of each event of $f^{(U)}$ appears in $f^{(C)}$, the bandwidth of the target user can affect the burst extraction process. As explained earlier in this section, we merge bursts of packets for messages whose IMD is less than $t_{e}$. Hence, suppose that the time it takes for the user to send a message is large enough to make the IMD between the current message and the next one less than $t_{e}$. Therefore, these two consecutive messages are combined in one burst. Table~\ref{bandwidth-p1} shows the value of $p_{1}$ observed from our data for different bandwidths. Since the bandwidth of our experiments is 1Mbps, $p_{1} = 0.921$.
{\tiny
\begin{table}[!t]
\caption{\small{The empirical value of $p_{1}$ measured for different client bandwidths}}
\centering
\begin{tabularx}{0.7\columnwidth}{ |Y|Y| }
	\hline
	Client Bandwidth (Mbps) & $p_{1}$ \\
	\hline
	0.1 & 0.824 \\
	\hline
	0.5 & 0.902 \\
	\hline
	1 & 0.921 \\
	\hline
	10 & 0.974 \\
	\hline
	100 & 0.983 \\
	\hline
\end{tabularx}
\label{bandwidth-p1}
\end{table}
}

Note that Type II error occurs when less than $\eta\cdot n$ messages of $f^{(C)}$ match $f^{(U)}$ when $H_{1}$ is true. Therefore,
\begin{align}
\nonumber \mathbb{P}_{\mathrm{FN}} = \mathbb{P}\left( k\leq \eta n |H_1\right) &= F(\eta n;n,p_{1})\\
\label{fn} & \leq \left(\frac {\eta}{p_{1}}\right)^{-n \eta } \left(\frac{1-\eta}{1-p_{1}}\right)^{\eta n-n},
\end{align}
where the last step follows from~\eqref{eq:1}.

We will validate the upper-bounds of $\mathbb{P}_{\mathrm{FP}}$ and $\mathbb{P}_{\mathrm{FN}}$ with our live experiments (Section~\ref{result:realworld}) and simulation results (Section~\ref{simulation}).

\subsection{Shape-Based Detector}\label{shape-based}

%

\begin{figure}[!t]
     \centering
     \includegraphics[width = \linewidth]{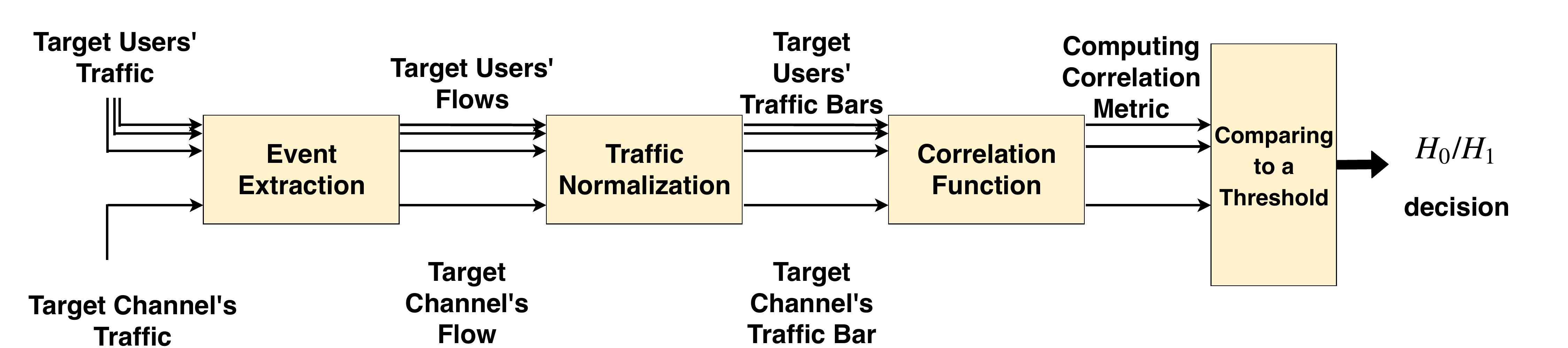}
     \caption{Shape-based detector}
     \label{fig:norm-corr-model}
\end{figure}

We design a second detector called the \emph{shape-based} detector. This detector links users to \im communications by correlating the shape of their network traffic, where traffic shape refers to the vector of packet lengths over time.
Figure~\ref{fig:norm-corr-model} illustrates the four stages of the shape-based detector.

\paragraphb{Event Extraction:} The first stage of the shape-based detector is to extract \im events from network traffic, which is performed similar to what was described earlier for the event-based detector.
As described in the following, we do this in a way that accounts for the different bandwidths of the users being correlated.


\paragraphb{Normalizing Traffic Shapes:}
The shape-based detector converts the extracted events into normalized traffic shapes by replacing each event with a traffic bar.
The reason for doing so is that the shape of an IM event (e.g., the corresponding packet burst) is a function of user network bandwidths; our traffic normalization removes the impact of user bandwidth, and therefore the adversary can correlate traffic shapes with no knowledge of the underlying users' bandwidths.

To perform this normalization, we replace each event (i.e., each burst) with a traffic bar whose width is $2\times t_{e}$, where
$t_{e}$ is the threshold used during event extraction as discussed in section~\ref{event-based}. We choose this value  to reduce the chances of overlaps between consecutive bars.
To capture the sizes of events in traffic normalization, the height of each bar is chosen such that the area under the bar equals  the size of the event.
Our shape normalization also reduces correlation noise by removing small traffic packets that are not part of any \im events.

To form the new normalized shape of traffic, we divide each bar into smaller bins of width $t_{s}$, the value of which is discussed in Section~\ref{param}, and with a height equal to the height of the corresponding bar. Therefore each bar consists of a number of bins of equal width and height. Furthermore, we put bins with the same width $t_{s}$ and height 0 between these bars. By doing so, after the traffic normalization, the new shape of traffic will be the vector of heights of bins over time.

\paragraphb{Correlating Normalized Traffic Shapes:}
Our shape-based detector correlates the normalized shapes of two traffic streams of target channel $C$ and user $U$ to decide if they are associated.
Suppose that $b^{(C)} = \{b^{(C)}_1,b^{(C)}_2,\ldots,b^{(C)}_{n_C}\}$ and $b^{(U)} = \{b^{(U)}_1,b^{(U)}_2,\ldots,b^{(U)}_{n_U}\}$ are the respective vectors of heights of bins associated with the target channel and user being tested, where $n_C$ and $n_U$ are the number of events in target channel and user flows, respectively.
We use the following normalized correlation metric:
\begin{equation}\label{cosine}
    corr = 2 \times \frac{\sum_{i=1}^{n}b^{(C)}_{i}b^{(U)}_{i}}{\sum_{i=1}^{n}(b^{(C)}_{i})^2 + \sum_{i=1}^{n}(b^{(U)}_{i})^2}
\end{equation}
where $n=\min(n_C,n_U)$. Note that $corr$ returns a value between 0 and 1, which shows the similarity of the two traffic shapes (1 shows the highest similarity).
Finally, the detector makes its decision by comparing  $corr$ to a threshold,  $	corr\underset{H_{0}}{\overset{H_{1}}{\gtrless}}\eta $,
where $\eta$ is the detection threshold.

\section{Attack Experiments}\label{exp-eval}

\subsection{General Setup}\label{setup}
We design our experimental setup to perform our attacks in the setting of Figure~\ref{fig:attack-model},
and based on the  threat model of Section~\ref{attack-model}.
We use the first type of  ground-truth in Figure~\ref{fig:attack-model} (adversary joins the target channel as a reading-only member), which is more challenging  for the adversary  compared to the  second ground-truth, and similar in performance compared to the third ground-truth mechanism.
Specifically, we use two \im clients using different \im accounts (e.g., Telegram accounts) that are running IM software on two separate  machines. One of these IM clients is run by the adversary, and the other one represents the target client.
The adversary client joins target channel $C$, (e.g., a public political \tg channel) and records the metadata of all the \im communications of $C$, i.e., the timing and sizes of all messages sent on that channel.
The target client  may or may not be a member/admin of the target channel $C$.
The adversary is not able to see the contents of the target client's communications (due to encryption), however she can capture the encrypted traffic of the target client. The adversary then uses the detection algorithms introduced in Section~\ref{detection-algo} to decide if the target user is associated with the target channel $C$.
In a real-world setting, the adversary will possibly have multiple target channels, and will monitor a large number of suspected clients.


For our \tg and Signal experiments, our adversary-controlled client uses their APIs to record \im communications of target channels, while for WhatsApp, we  manually send messages through its Desktop version (as it does not have an API).



\paragraphb{Parameter Selection.}\label{param}
We choose burst detection threshold as $t_{e}=0.5s$ based on the empirical distribution of network jitter.  Also,  we set $t_{s}$ of the shape-based detector to $0.01s$, as it leaves enough separation between two consecutive IM messages.
We set $\Delta$ of the event-based algorithm to 3 seconds. 
We also set $\Gamma$ parameter of the event-based detector to $10Kb$.

\paragraphb{Ethics.}\label{sec:ethics}
We performed our inference  attacks only over public IM channels; therefore, we did not capture any private IM communications.
Also, we performed our attacks only on our own IM clients, but no real-world IM clients. Therefore, our experiments did not compromise the privacy of any real-world IM members or admins.

\subsection{In-The-Wild Attacks on \tg}\label{result:realworld}
We experiment our attacks on in-the-wild \tg channels (fully complying with the ethical considerations of Section~\ref{sec:ethics}).

\paragraphb{Experimented Channels}
As discussed in Section~\ref{data}, \im services put limits on the number of channels a client can join. For our experiments, we joined 500  channels popular among Iranian users, with different rates of daily messages.
In our experiments, each time our own client connects to one of these 500 channels, and the goal of the adversary is to match our client to the channel she has joined.

\paragraphb{Synchronization} As the adversary's clock may be skewed across her vantage points,
our adversary uses a simple sliding window to mitigate this: for the first 10 seconds of traffic, the adversary slides the two flows being compared with 0.5 second steps, and uses the maximum correlation value.

\begin{figure}[!t]
\begin{subfigure}{0.49\linewidth}
     \centering
     \includegraphics[width = 0.99\linewidth, trim = {1cm 0cm 1cm 1cm}]{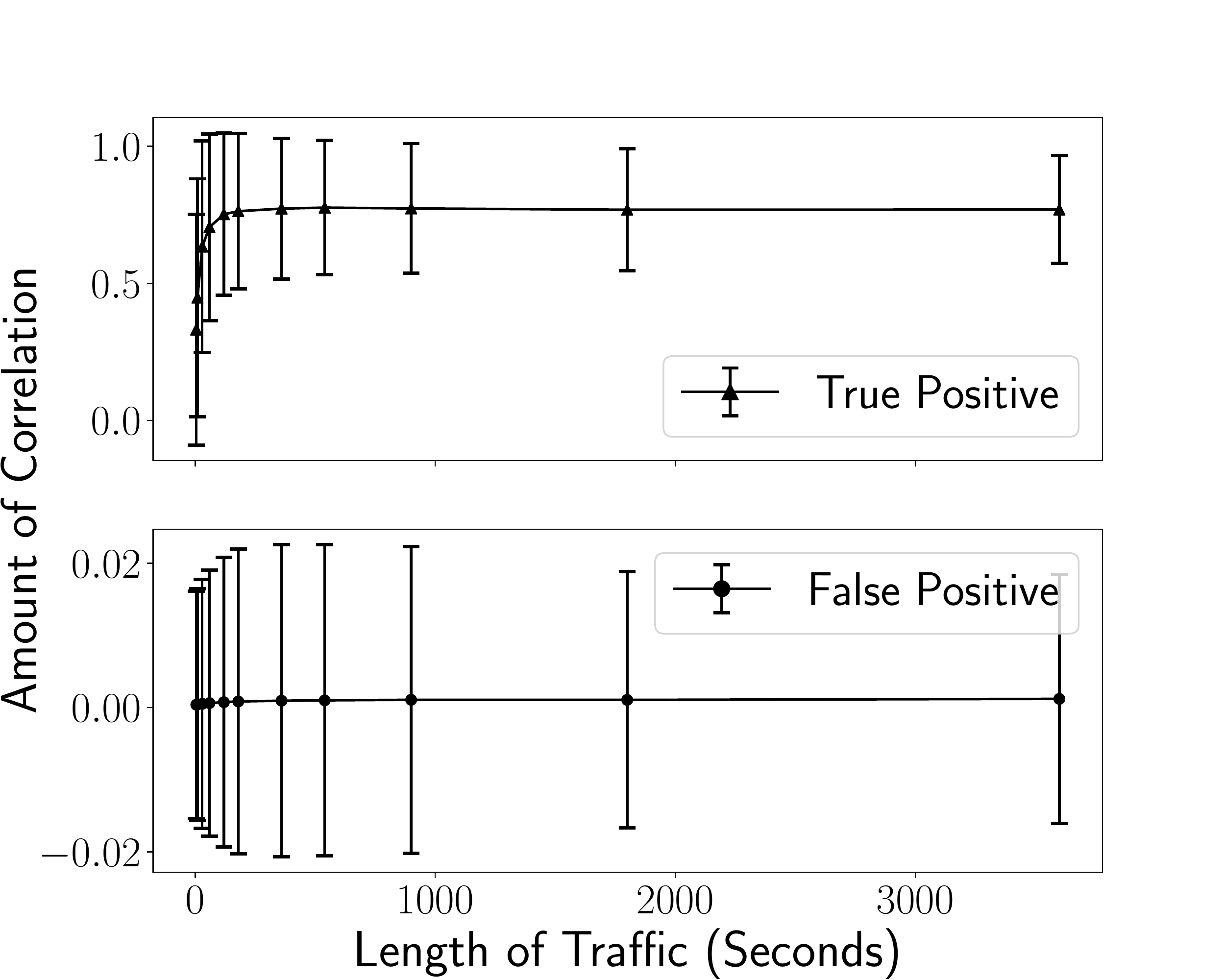}
     \caption{Shape-based detector}
\end{subfigure}
\begin{subfigure}{0.49\linewidth}
     \centering
     \includegraphics[width = 0.99\linewidth, trim = {1cm 0cm 1cm 1cm}]{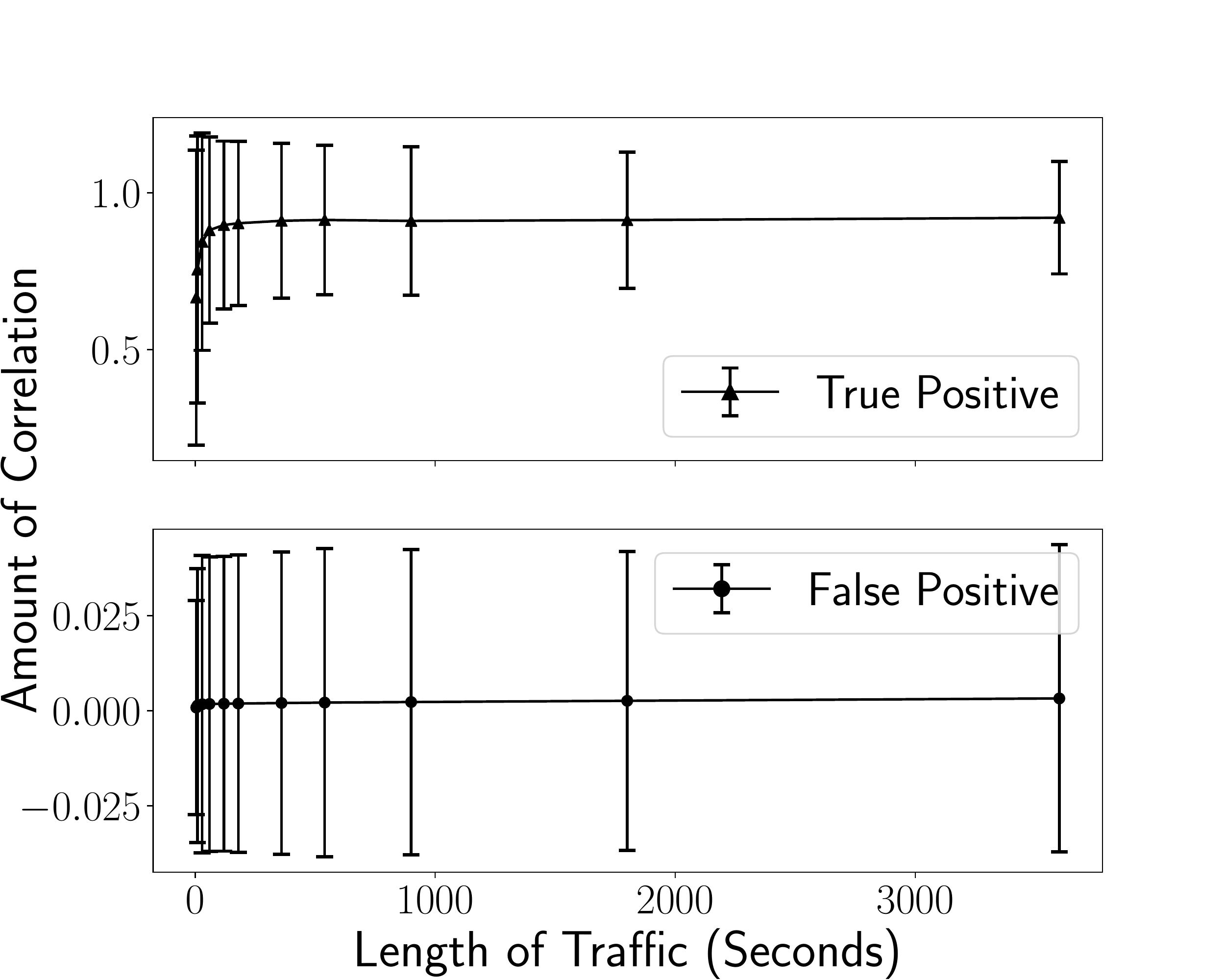}
     \caption{Event-based detector}
\end{subfigure}
\caption{Attack correlation metrics for different traffic lengths. The  performance improves for longer observations.}\vspace{-3ex}
\label{fig:actual-corr-per-length}
\end{figure}

\paragraphb{Choosing the Threshold}
Figure \ref{fig:actual-corr-per-length} shows the TP  and FP rates of our experiments for different detection thresholds $\eta$, and for different traffic lengths (traffic length excludes the long inactivity periods across correlated connections).
Each point in the graph shows the average of the correlation metric across all experiments, and the bars show the standard deviation.
Intuitively, the detector performs well for wider gaps between the TP and FP bars.
From the figures, we see the impact of the traffic length on detection performance: \emph{longer observations improve detection performance} of our attacks.
Also,  $\eta$ trades off the TP and FP rates.
The adversary can detect the right threshold  and the right traffic length based on her target TP and FP values.

\begin{figure*}[!htb]
\minipage{0.3\textwidth}
  \includegraphics[width = \linewidth, trim = {2cm 0cm 2cm 2cm}]{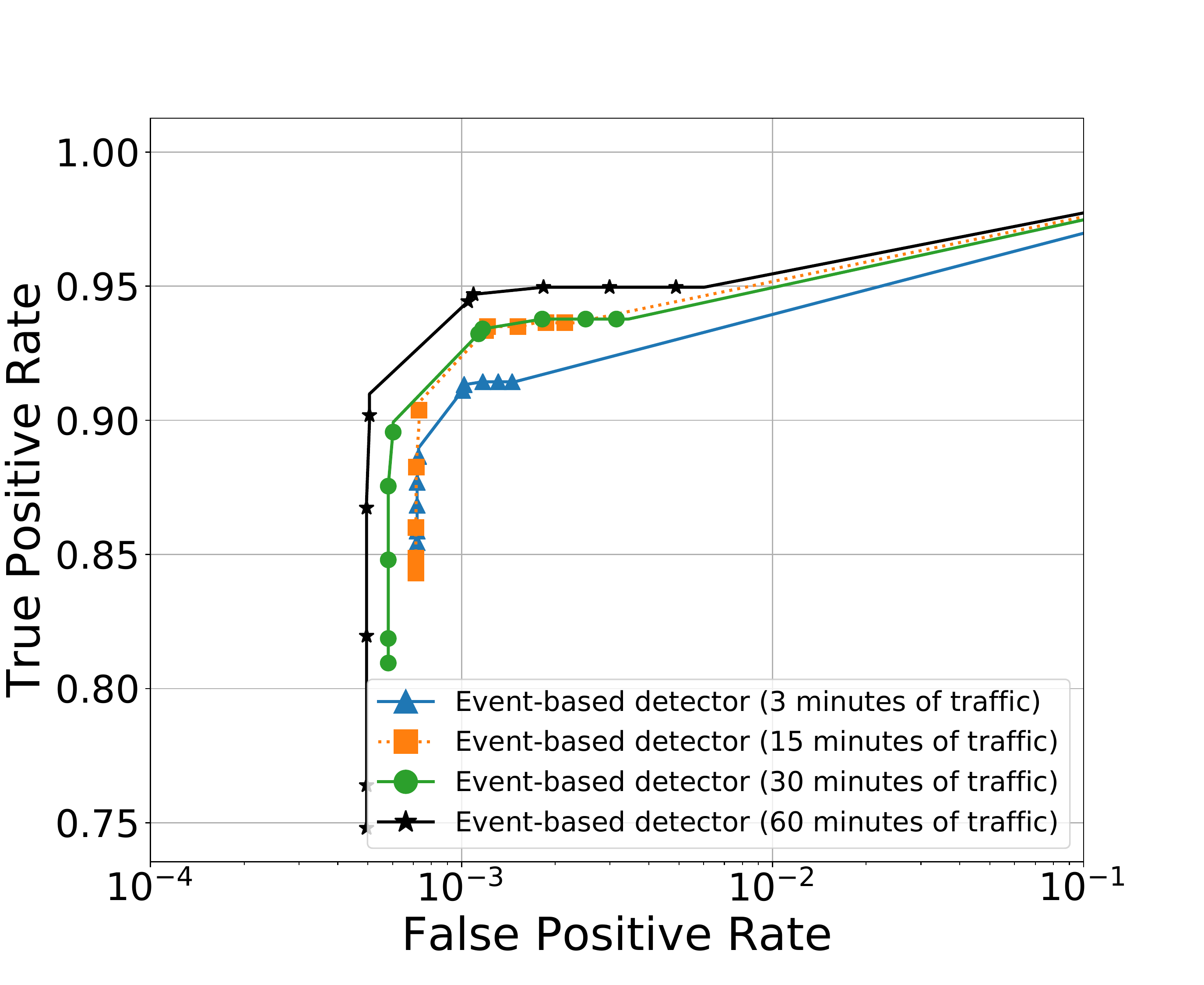}
     \caption{Performance of the event-based detector over in-the-wild \tg traffic}
     \label{fig:roc-actual-size}
\endminipage\hfill
\minipage{0.3\textwidth}
  \includegraphics[width = \linewidth, trim = {2cm 0cm 2cm 2cm}]{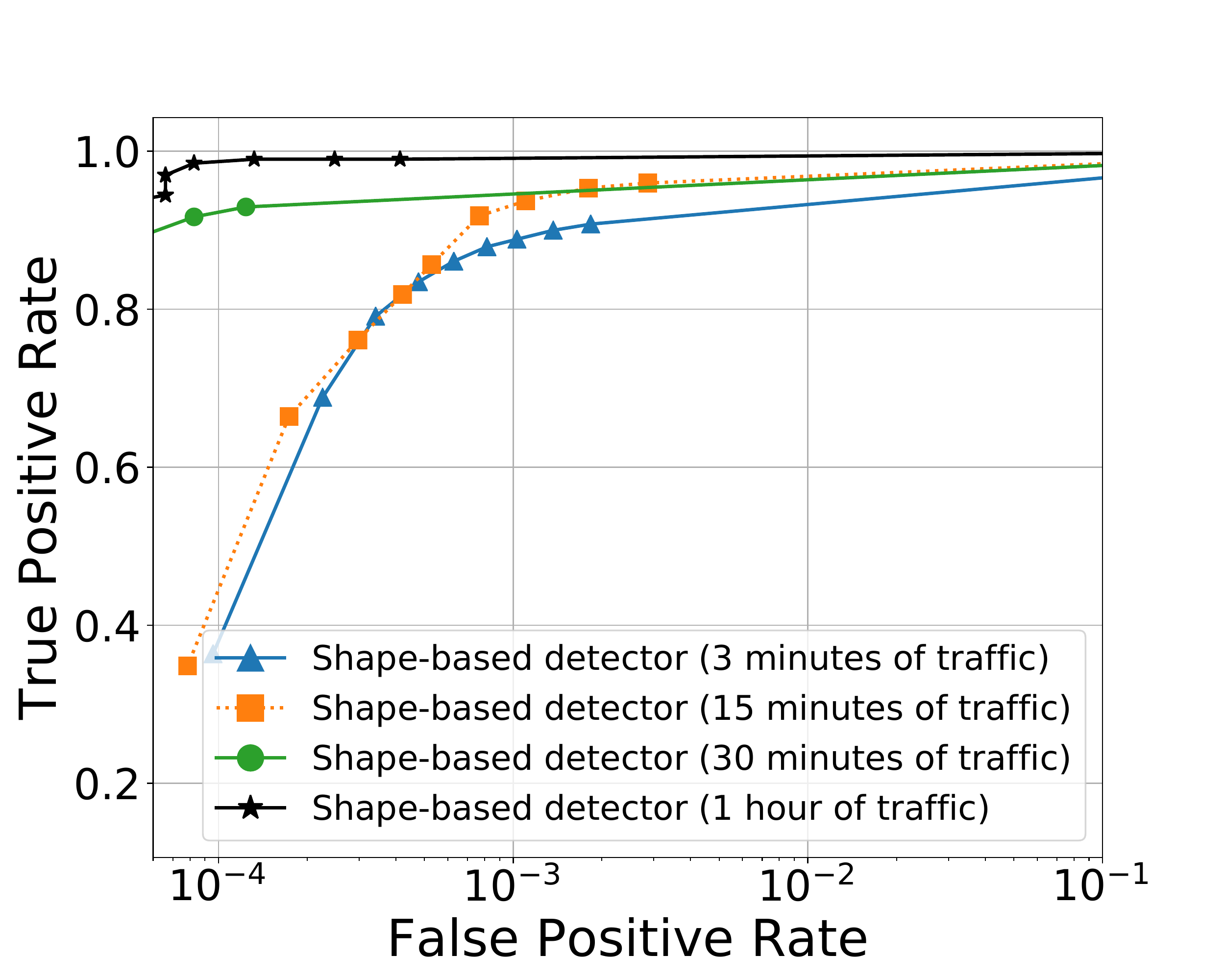}
     \caption{Performance of the shape-based detector over in-the-wild \tg traffic}
     \label{fig:roc-actual-cosine}
\endminipage\hfill
\minipage{0.3\textwidth}
  \includegraphics[width = \linewidth, trim = {2cm 0cm 2cm 2cm}]{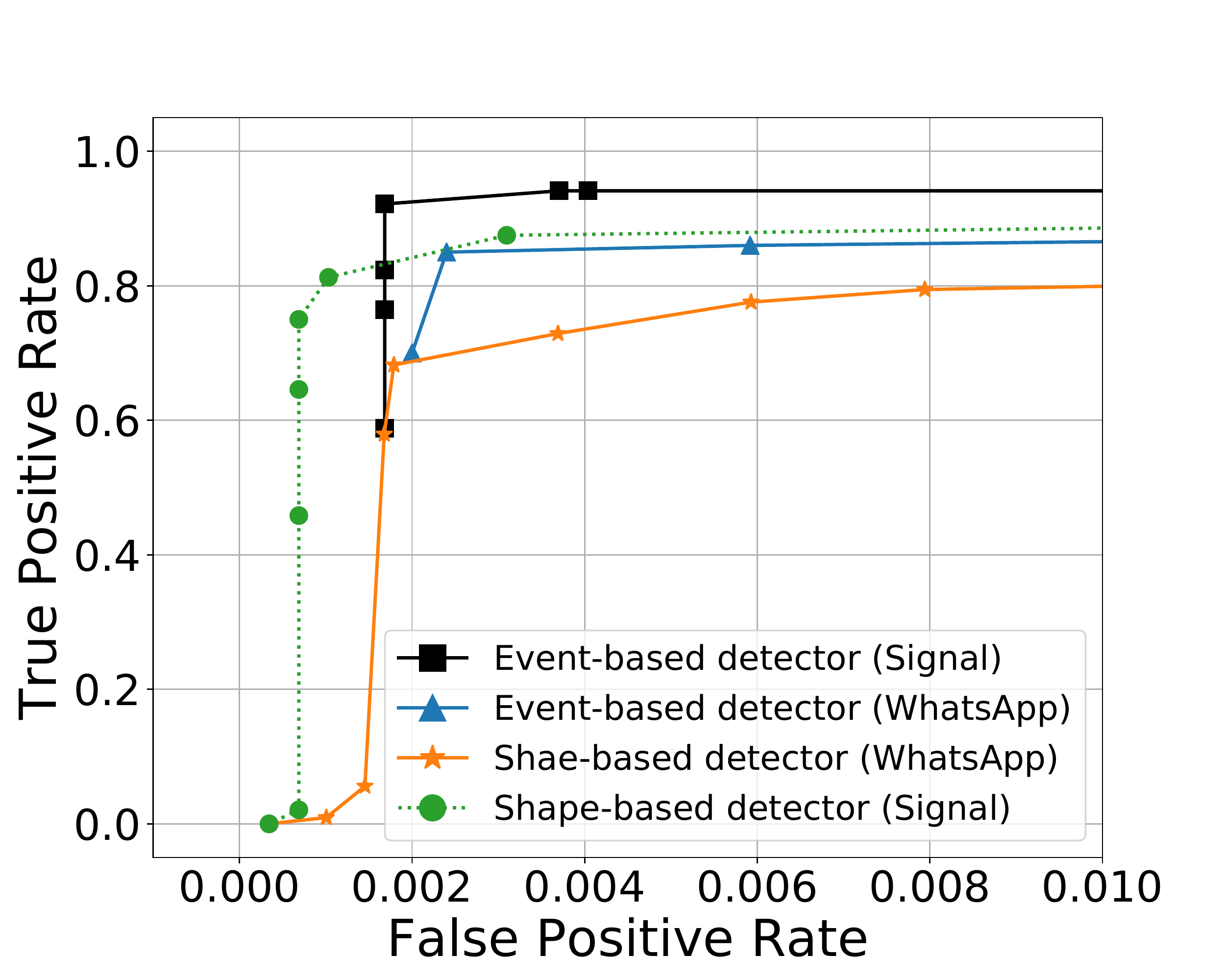}
     \caption{Performance of event-based and shape-based detector on Signal and WhatsApp}
     \label{fig:roc-signal-whatsapp}
\endminipage
\end{figure*}

\paragraphb{Comparing our Two Attacks:}
Figures~\ref{fig:roc-actual-size} and \ref{fig:roc-actual-cosine} show the performance of event-based and shape-based detectors, respectively, for different detection thresholds using an ROC curve (for 4 different observation lengths). We can see that, as expected,  \emph{longer traffic observations improve the accuracy of both detectors}.
For instance, the shape-based detector offers a TP~$=.94$ and FP~$=10^{-3}$ with 15 min observation, while an hour of observation reduces the FP to close to FP~$=5\times 10^{-5}$.
In practice, \emph{an adversary can deploy the attack with hierarchical observation intervals to optimize accuracy and computation}. For instance, the adversary can monitor a mass of IM users for 15 mins of observation; then the adversary will  monitor only the clients detected with 15 mins observations for  longer time periods, e.g., an hour, to improve the overall  FP performance while keeping computations low.

Furthermore, as can be seen, \emph{the shape-based detector outperforms the event-based detector for smaller values of false positive rates}. For instance, for a target true positive rate of 0.9, the shape-based detector offers a false positive of $5\times 10^{-4}$ compared to $8\times 10^{-4}$ of the event-based detector (with 15 mins of observation).
The performance gap decreases for higher false positive rates.
The reason for this performance gap is the impact of event extraction noise on the event-based detector. Such noise has smaller impact on the shape-based detector as it correlates the shape of traffic flows.


On the other hand, \emph{our event-based detector is two orders of magnitude faster than the shape-based detector}.
Table~\ref{method-time} compares the correlation times of the two detectors (averaged over 100 experiments).
The main reason for this difference is that the event-based correlator uses the discrete time-series of event metadata for its correlation, while the shape-based detector uses traffic histograms over time.
{
\tiny
\begin{table}
\centering
\caption{Comparing the runtimes of our two attacks.}
\resizebox{0.7\linewidth}{!}{
\begin{tabular}{ |c|c| }
	\hline
	\textbf{Method} & \textbf{One correlation time} \\
	\hline
	Shape-based correlation & 167$\mu s$ \\
	\hline
	Event-based correlation & 2$\mu s$ \\
	\hline
\end{tabular}}
\label{method-time}
\end{table}
}

Note that for our event-based detector in Figure~\ref{fig:roc-actual-size}, for short traffic observations (e.g., 15 mins)  we cannot observe small FPs in our ROC curve. This is because the event-based correlation  uses the number of matched events, which is very coarse-grained due to the limited number of events in short (e.g., 15 minutes) intervals. We use our analytical upper-bounds (derived in~\eqref{fp} and \eqref{fn}) to  estimate the performance trend for smaller false positive values (Figure~\ref{fig:fp-fn} of Appendix).



\subsection{In-The-Wild Attacks on WhatsApp and Signal}
As discussed previously in Section~\ref{data}, we make the bulk of our data collection and experiments on \tg due to its huge number of public channels (making our experiments ethical and realistic as we do not need to do experiments on private communications).
However, as shown in Figure~\ref{other-sims}, the shape of traffic is similar across different \im services, and therefore we expect our attack algorithms to perform similarly when applied by an adversary on other \im applications.
We validate this through experiments on Signal and WhatsApp messengers.

Signal and WhatsApp only offer private (closed) channels. For ethical reasons, we make our own (closed) channels on these services to perform our experiments. Specifically, we create a private channel on each of Signal and WhatsApp. We send messages on these channels by mimicking the patterns (i.e., inter-message times and message sizes) of randomly chosen public \tg channels. Our user and adversary VMs join these channels, and we perform our attacks in the same setting as the \tg experiments.

Figure~\ref{fig:roc-signal-whatsapp} shows the performance of event-based and shape-based detectors in both Signal and WhatsApp applications using 15 minutes of traffic, \emph{demonstrating  detection performances comparable to those of \tg}. In particular, similar to \tg, for smaller false positive rates, the shape-based detector has better performance while the event-based detector achieves more accuracy for larger false positive rates.
Our results show that \textbf{our attacks generalize to \im{s} other than \tg due to the similarity of their traffic patterns}, which is due to these services  not using any obfuscation mechanisms.


\subsection{Comparison with Deep Learning Techniques}\label{deepcorr}


As mentioned earlier in Section~\ref{related-work}, the recent work of  DeepCorr~\cite{Nasr:2018:DSF:3243734.3243824} uses  deep learning classifiers to perform flow correlation attacks on Tor. They demonstrate that deep learning classifiers outperform statistical correlation techniques, like the ones we used in our work, in correlating Tor connections.
In this section, we compare our IM classifiers with deep learning classifiers.
As we show in the following,  \emph{our statistical classifiers outperform  deep learning-based classifiers}, specially for shorter flow observations. Intuitively, this is due to the sparsity of  events in typical  IM communications, as well as the  stationary nature of noise in IM communications in contrast to the scenario of Tor. Note that this fully complies with Nasr et al.~\cite{Nasr:2018:DSF:3243734.3243824}'s observation that DeepCorr only  outperforms statistical  classifiers in non-stationary noisy conditions, where statistical traffic models become inaccurate.

For fair comparisons, we obtain the original code of DeepCorr~\cite{Nasr:2018:DSF:3243734.3243824},   and adjust  it to the specific setting of IM traffic.
Specifically, we divide the timing of each flow to equal periods of length $1$ second, and in each period we assign values of $\{0,1\}$ to that period. We set the value of a period $1$ if there is a burst of packets in that period, and $0$ if there is no burst of packets.
As an example, if we use 15 minutes of traffic flows for correlation, our feature dimension is a  $900$-length vector with values of $0, 1$.
We train our classifier using $500$ associated flow pairs and $2,000$ non-associated flow pairs. We test our (DeepCorr-based and statistical) classifiers   using a non-overlapping set of $200$ associated flow pairs and $4,000$ non-associated.

\begin{figure}[]
     \centering
     \includegraphics[width = 0.8\linewidth, trim = {1cm 1cm 1cm 1cm}]{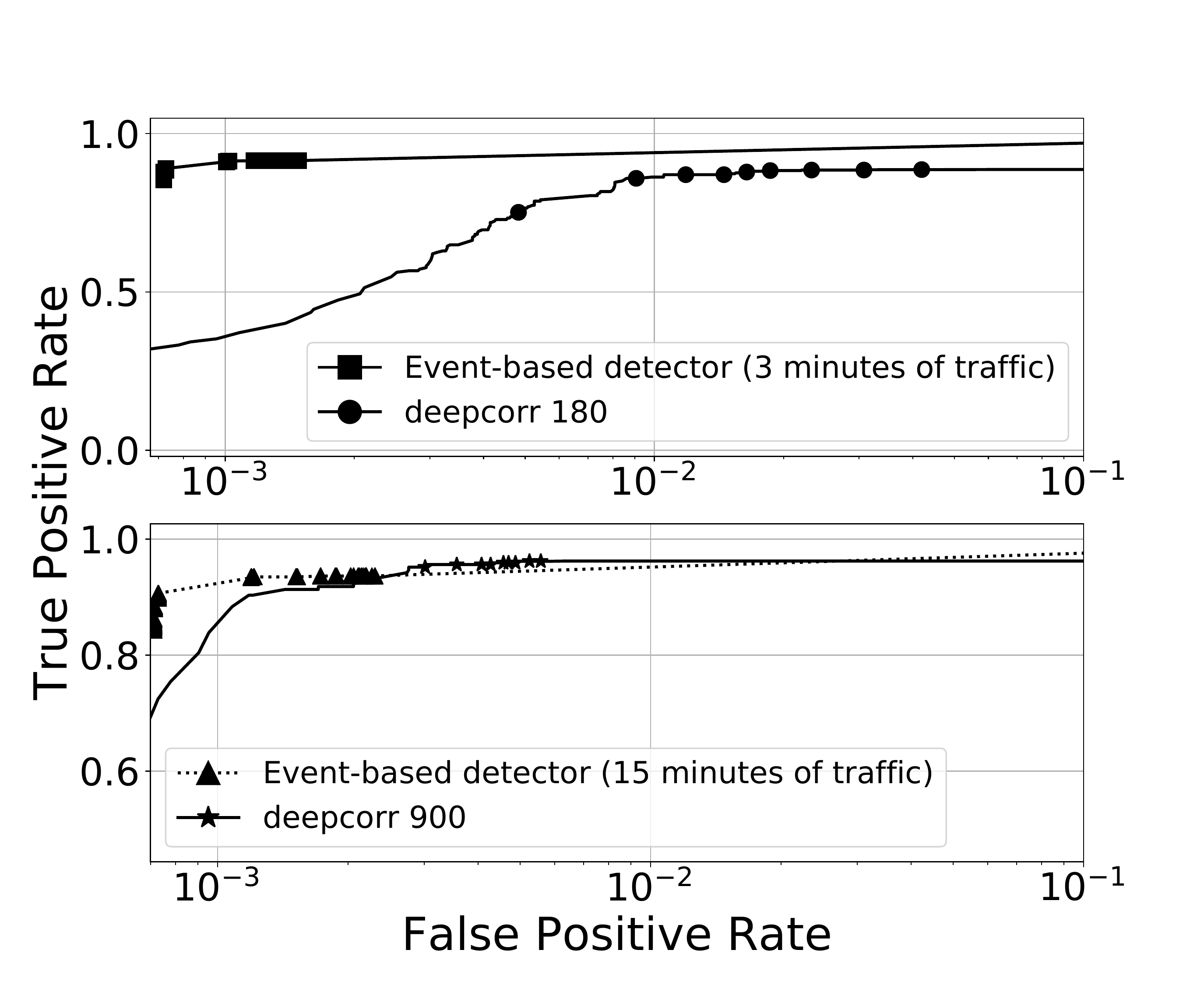}
     \caption{Comparing our event-based detector with a  DeepCorr-based classifier, for 3 and 15 minutes of traffic observation}\vspace{-3ex}
     \label{fig:deepcorr-comparison}
\end{figure}

Figure~\ref{fig:deepcorr-comparison} shows the ROC curves of our event-based detector compared with our deep learning-based detector, using $3$ and $15$ minutes of traffic. As we can see, \emph{our event-based technique outperforms the  deep learning-based classifier for smaller false positive rates}. For instance, for a false positive rate of $10^{-3}$ and using 15 minutes of traffic, our event-based detector achieves a $93\%$ accuracy compared to $88\%$ of the  DeepCorr-based technique.
We see that the  performance advantage of our event-based detector significantly increases for shorter flow observations, e.g., when 3 minutes of traffic is used for detection, our classifier provides  $92\%$ accuracy compared to  $45\%$ of the DeepCorr-based classifier (for the a false positive rate of $10^{-3}$).

\subsection{Simulations Using Synthetic Traffic}\label{simulation}

Our in-the-wild experiments, presented above, have been done on a limited number of \im channels, which is due to the fact that major \im services put limits on the number of channels a client can join.
To ensure the reliability of our results, we generated a large number of synthetic \im channels (using the algorithm from Section~\ref{synthesize}), and evaluated our attack performance on them.
\emph{Our evaluations using 10,000 synthetic IM channels complies with our results from our in-the-wild experiments}. We have presented our synthetic evaluations in Appendix~\ref{appendix:simulation}.

%
%
%

\subsection{Discussions}
\subsubsection{Impact of Other Channels}
In our experiments, a target user is the member/admin of only a single target channel, while in practice a user may be a member/admin of multiple channels.
Therefore, a valid question is whether the traffic patterns of other channels may interleave with the patterns of the target channel and reduce the reliability of our detectors.
We argue that this will not be an issue as long as the detection is performed during the time interval the user is visiting/posting to the target channel.
This is because when a user is visiting a target channel, he will not receive the messages sent to other channels (he will only receive some small-sized notifications). Also, if an admin user simultaneously sends a message to multiple channels, his upstream traffic (to the IM server) will only contain a single message.
Therefore, to identify whether a given user is the member/admin of a target channel, the adversary needs to continuously monitor that user until the user visits or posts to   the target channel.


\subsubsection{Impact of Network Conditions}
While we have performed our experiments in specific network conditions, we argue that our detectors will perform equivalently in other network conditions as well.
This is because in our presented threat model, the adversary has knowledge on the network conditions of each target user (e.g., the adversary can be the ISP of the target user), and therefore she can adjust the detector for various users, as shown in Table~\ref{bandwidth-p1}. Also, note that natural variations in a user's network conditions will not  impact the detectors as IM traffic patterns are  resilient to  natural network perturbations (e.g., the distance between IM bursts as shown in Figure~\ref{other-sims} are orders of magnitude larger than network jitter).

\section{Countermeasures}

\begin{figure*}[!htb]
\minipage{0.3\textwidth}
  \includegraphics[width = \linewidth, trim = {2cm 0cm 2cm 2cm}]{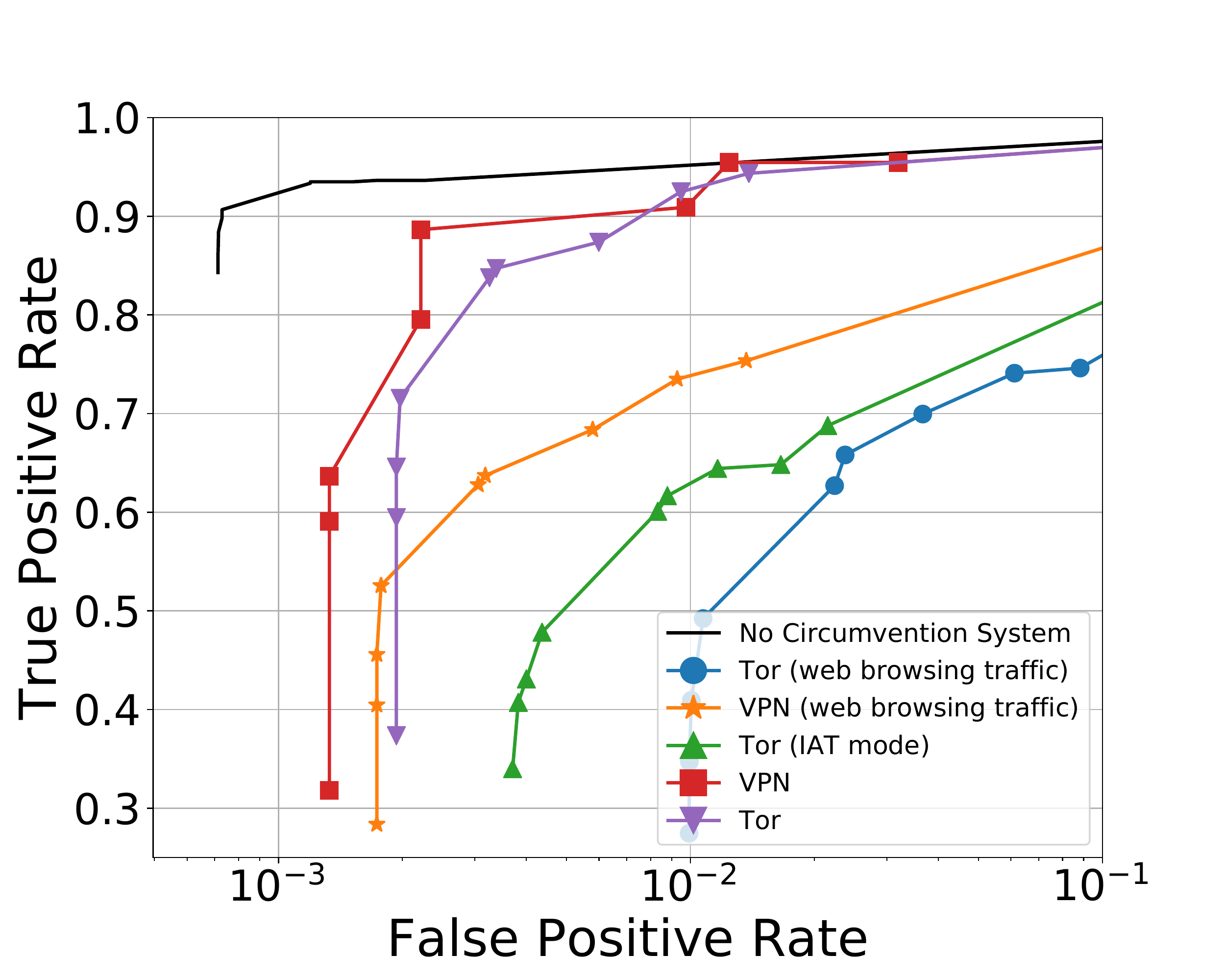}
     \caption{The impact of various countermeasures on the performance of the event-based detector using different circumvention systems (15 minutes of observed traffic)}
     \label{fig:roc-cm}
\endminipage\hfill
\minipage{0.3\textwidth}
  \includegraphics[width = \linewidth, trim = {2cm 0cm 2cm 2cm}]{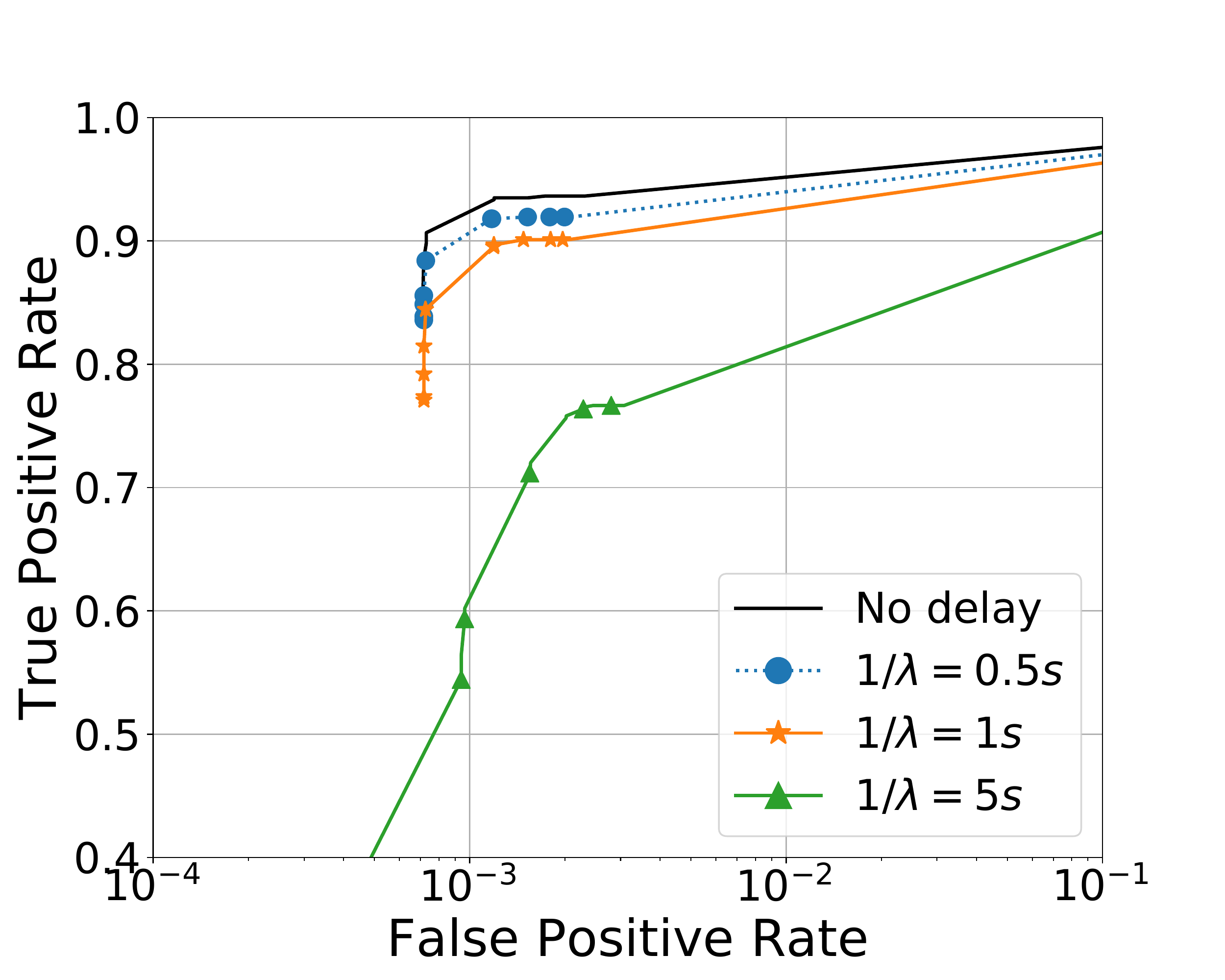}
     \caption{Randomly delaying events by an \im server acts as an effective countermeasure to our attacks. $\frac{1}{\lambda}$ is the mean of the added delay (15 minutes of observed traffic)}
     \label{fig:roc-delay}
\endminipage\hfill
\minipage{0.3\textwidth}%
  \includegraphics[width = \linewidth, trim = {2cm 0cm 2cm 2cm}]{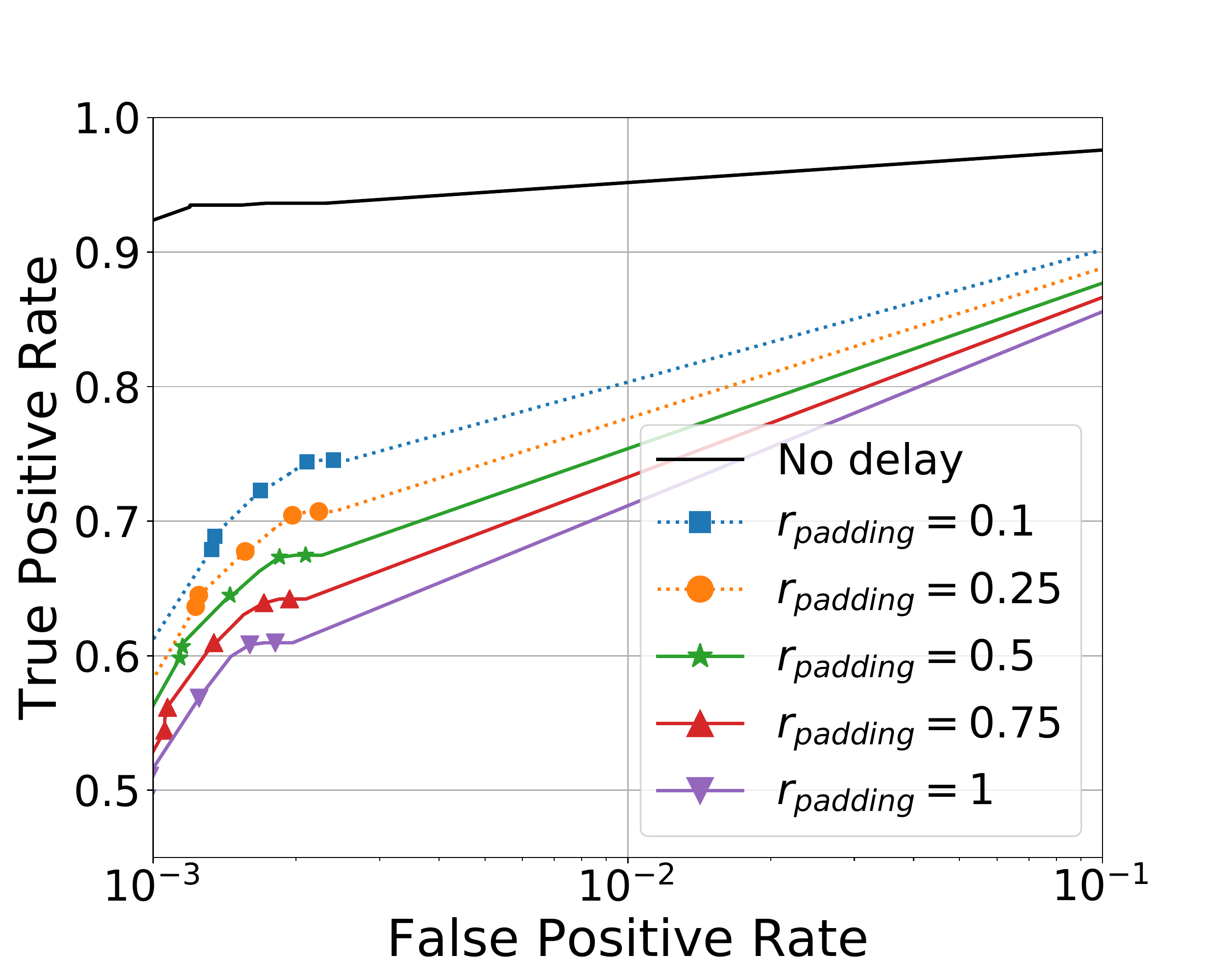}
     \caption{Padding IM events by the \im server (or client) can act as an effective countermeasure against our attacks.  (15 minutes of observed traffic)}
     \label{fig:roc-cover}
\endminipage
\end{figure*}

We deploy and evaluate possible countermeasures against our  presented attacks.
Intuitively, our attacks work because in-the-wild \im services do not deploy any mechanisms to obfuscate traffic patterns.
Therefore, we investigate various traffic obfuscation mechanisms as countermeasures against our traffic analysis-based attacks.

Note that obfuscation-based countermeasures have been studied against other kinds of traffic analysis attacks overviewed in Section~\ref{related-work}. There are several key ideas used in  existing countermeasures:
(1) tunneling traffic through an overlay system that perturbs its  patterns~\cite{obfs4, MohajeriMoghaddam:2012:SPO:2382196.2382210}, e.g., Tor, (2) adding background traffic (also called decoy) that is mixed with the target traffic~\cite{Dyer:2012:PIS:2310656.2310689, morphing, Luo11httpos:sealing, Panchenko:2011:WFO:2046556.2046570, Wang2017WalkieTalkieAE}, (3) padding traffic events (e.g., packets)~\cite{fingerprinting-esorics2016, Cai:2014:CCS:2665943.2665949, Cai:2014:SAD:2660267.2660362, Dyer:2012:PIS:2310656.2310689, Wang:2014:EAP:2671225.2671235}, and (4) delaying traffic events~\cite{Cai:2014:CCS:2665943.2665949, Cai:2014:SAD:2660267.2660362, Dyer:2012:PIS:2310656.2310689, Wang:2014:EAP:2671225.2671235, Wang2017WalkieTalkieAE}.
In the following, we investigate various countermeasure techniques inspired by  these standard approaches.



\subsection{Tunneling Through Circumvention Systems With/Without Background Traffic}\label{tunnel}

As the first countermeasure, we tunnel \im traffic through standard circumvention systems, in particular  VPN and Tor pluggable transports~\cite{tor-pt}. We use the same experimental setup as before and connect to 300 Telegram channels. For each circumvention system, we  perform the experiments with and without any  background traffic.
In the experiments with background traffic, the VM running the \im software also makes HTTP connections using Selenium. The background HTTP webpages are picked randomly from the top 50,000 Alexa websites. To amplify the impact of the background traffic, the time between every two consecutive HTTP GETs is taken from the empirical distribution of \tg IMDs, therefore producing a noise pattern similar to actual \im channels.

\emph{We observe that our event-based attack performs stronger against our countermeasures}. Therefore,
we only present the countermeasure results against the event-based detector (we show the results for the shape-based attack in Appendix~\ref{cm-cosine}).
Figure~\ref{fig:roc-cm} shows the ROC curve of the event-based detector using various circumvention systems and in different settings.  Our Tor experiments are done once with regular Tor, and once using the obfs4~\cite{obfs4} transport with the IAT mode of 1, which obfuscates traffic patterns.

We see that \emph{using regular Tor (with no additional obfuscation) as well as using VPN does not significantly counter our attacks}, e.g., we get a TP of $85\%$ and a FP of $5\times 10^{-3}$ when tunneling through these services (using 15 mins of traffic).
However, adding \emph{background traffic} when tunneled through Tor and VPN reduces the accuracy of the attack, but \emph{we get the best countermeasure performance using Tor's obfs4 obfuscator}.

Note that tunneling through a generic circumvention system like Tor is not the most attractive countermeasure to the users due to the poor connection performance of such systems.

\subsection{\impr: An Obfuscation Proxy Designed for IM Services}\label{obfs}
We design and implement a proxy-based obfuscation system, called \impr\footnote{\url{https://github.com/SPIN-UMass/IMProxy}}, built specifically for IM communications.
\impr combines two obfuscation techniques:
 changing the timing of events (by adding delays), and changing the sizes of events through adding dummy traffic.
 An IM client has the ability to enable each of these countermeasures, and specify the amplitude of obfuscation to make her desired tradeoff between performance and resilience.
 \emph{\impr does not require any cooperation from IM providers}, and can be used to obfuscate \emph{any IM service}.

\begin{figure}[!t]
     \centering
     \includegraphics[width = 1\linewidth]{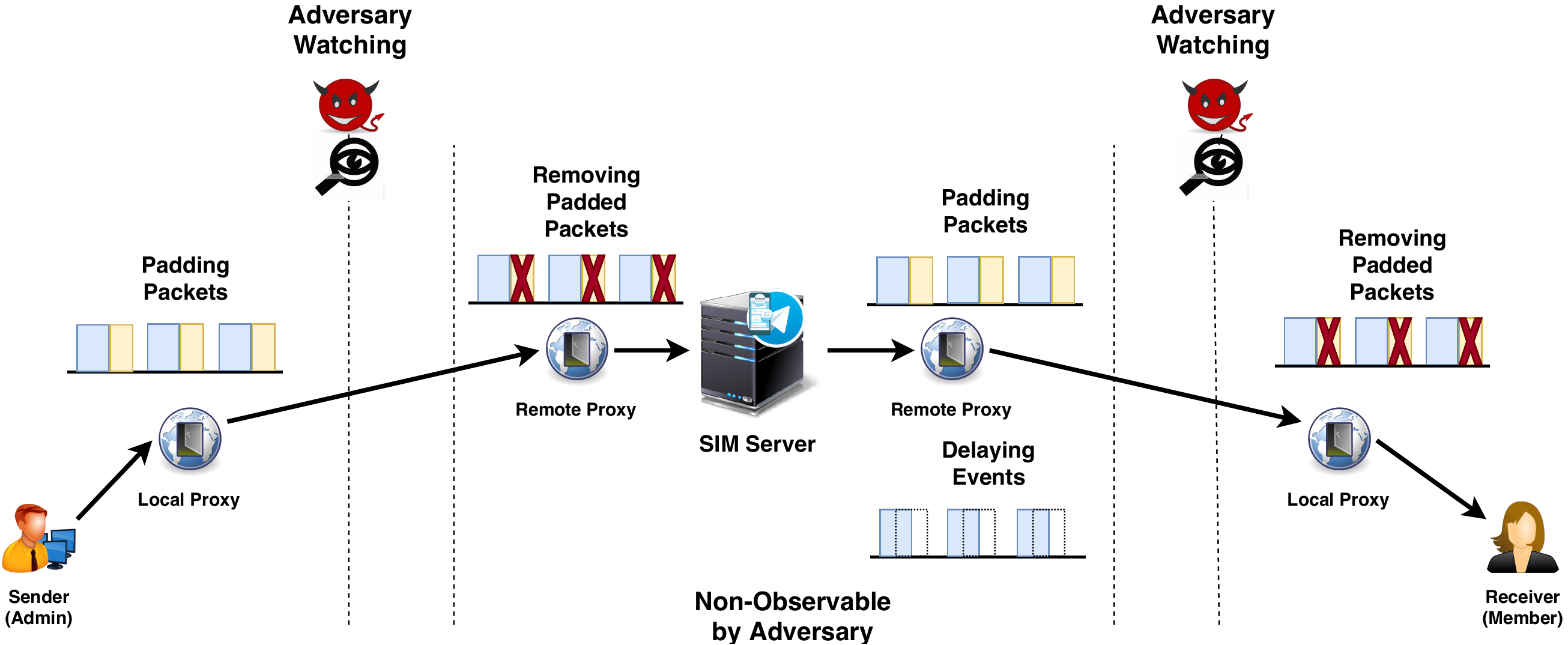}
     \caption{Design of our \impr countermeasure.}
     \label{fig:proxy-model}
\end{figure}

\paragraphb{Components of \impr:}
Figure~\ref{fig:proxy-model} shows the design of \impr.
For a client to use \impr, she needs to install a Local\impr software. Local\impr runs a  SOCKS5 proxy listening on a local port. The client  will need to change the setting of her IM software (e.g., Telegram software) to use this local port for proxying.

A second component of \impr is Remote\impr, which is a SOCKS5 proxy residing outside of the surveillance area. The client needs to enter the (IP,port) information of this remote proxy in the settings of her Local\impr software.  Note that, in practice, Remote\impr can be  either run by the client herself (e.g., as an AWS instance),  or can be run by the IM provider or trusted entities (similar to the MTProto proxies run for Telegram users~\cite{mtproto-proxy}).

\paragraphb{How \impr works:}
Once an IM client sets up her system to use \impr as above, her IM traffic to/from the IM servers will be proxied by the proxies of \impr, as shown in Figure~\ref{fig:proxy-model}. The IM traffic of the client will be handled  by Local\impr and Remote\impr, which obfuscate traffic through padding and delaying.

As shown in the figure, \impr acts differently on upstream and downstream IM traffic.
For upstream \im communications (e.g., messages sent by an admin), Local\impr adds padding to the traffic by injecting dummy packets and events at certain locations.
First, some dummy packets are injected close to the  events in order to change their sizes. The size of padding for each event is chosen randomly, following a uniform distribution in [0,$r_{padding}$], where  $r_{padding}$ is a parameter adjusted by each user.
Second, some dummy events (burst of packets) are injected during the silence intervals; this is done randomly: during each 1 second silence interval, an event is injected with a probability $p_{padding}$, where $p_{padding}$ is also adjusted by each individual user.
The size of dummy events is drawn from the empirical distribution of the sizes of image messages, as presented earlier.
Finally, the dummy packets are removed by Remote\impr before getting forwarded  to the IM server. Note that all traffic between Local\impr and Remote\impr is encrypted so the adversary can not identify the dummy packets.

For downstream \im communications (e.g., messages received by a member), Remote\impr adds dummy packets, as above, which are dropped by Local\impr before being released  to the client's IM software.
In addition to padding, Remote\impr delays the packets in the downstream traffic.
In our implementation, Remote\impr uses an  Exponential Distribution  with  rate $\lambda$ to generate random delays (which is based on our delay model in Figure~\ref{fig:qq-delay}).
Note that no delay is applied on upstream traffic, as the delay will transit to the corresponding downstream traffic.

Note that each  client can control the intensity of padding by adjusting the $p_{padding}$ and  $r_{padding}$ parameters, and control
 control the amplitude of  delays by adjusting  $\lambda$.



\paragraphb{Implementation:}
We have implemented \impr in  Python using the \textit{socketserver} module. We use a Threading TCP Server and Stream Request Handler to implement the SOCKS5 proxy in python.
We have released our software as open source.

\paragraphb{Evaluation against oblivious adversary:}
We first evaluate our \impr implementation against an adversary who is not aware of how \impr works (or its existence). To do so, we evaluate \impr against our event-based detector.

Figure~\ref{fig:roc-delay} shows the ROC curve of the event-based detector for different values of $\lambda$. Note that $\frac{1}{\lambda}$ defines the average amount of delay added to the packets.
As we can see, \emph{increasing the added delay (by reducing $\lambda$) reduces the performance of our attack}, as it causes to missalign events across the monitored flows. For instance, a $\frac{1}{\lambda} = 1s$ reduces the adversary's TP from $93\%$ to $86\%$ (for a constant $10^{-3}$ false positive).

Figure~\ref{fig:roc-cover} shows the ROC curves of the event-based detector with different $r_{padding}$ and $p_{padding} = 10^{-4}$.
Note that a $p_{padding} = 10^{-4}$ causes a $7\%$ average traffic overhead (please refer to Appendix~\ref{padding} on how  bandwidth overhead is calculated).
As expected, \emph{increasing $r_{padding}$ reduces the performance of our attack}; even a $r_{padding}$ as small of $10\%$ and $7\%$ of dummy events can have a noticeable impact on countering the traffic analysis attacks, i.e., for a  $10^{-3}$ false positive rate, the detection accuracy is reduced from $93\%$ to $62\%$. Increasing $r_{padding}$ to $50\%$ will further reduce detection  accuracy to $56\%$.

%
%

\paragraphb{Evaluation against \impr-aware adversary:} Next, we evaluate \impr against an adversary who is aware that target users are deploying \impr and also knows the details of \impr.
Our adversary trains a DeepCorr-based classifier on IM traffic obfuscated using \impr (note that our statistical detectors will suffer for such an adversary due to the randomness of \impr's obfuscation).

\begin{figure}[t]
     \centering
     \includegraphics[width = 0.8\linewidth, trim = {1.2cm 1.2cm 1.2cm 1.2cm}]{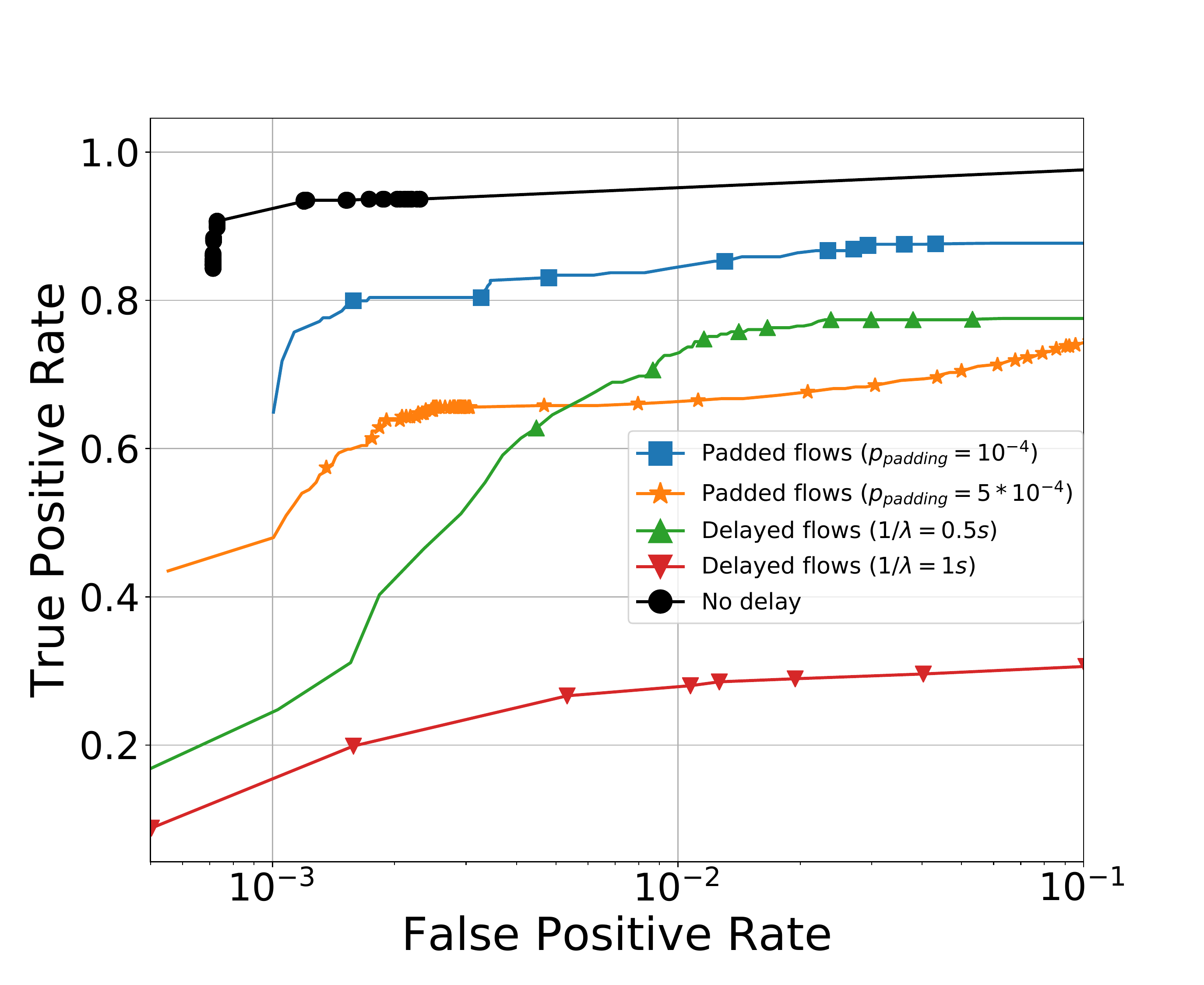}
     \caption{Evaluating \impr against an \impr-aware classifier (trained using DeepCorr).}\vspace{-3ex}
     \label{fig:deepcorr-delay}
\end{figure}

Figure~\ref{fig:deepcorr-delay} shows the performance of this DeepCorr-based classifier against \impr-obfuscated connections (each flow is 15 mins).
We use  $r_{padding}=0.1$  and evaluate the performance for different values of $p_{padding}$.
As can be seen,  \emph{\impr is highly effective even against an \impr-aware classifier}, demonstrating   \impr's efficiency in manipulating IM traffic patterns.
For instance,
for a false positive rate of $10^{-3}$, the \impr-aware classifier provides true positive rates of $25\%$ and $15\%$ (for average obfuscation delays of $0.5$ and $1$), which is significantly weaker compared to $93\%$ of the  event-based detector when \impr is not deployed.
%
As we can see, delaying provides better protection than padding; however, we expect that most users will prefer padding over delays due to the latency-sensitive nature of IM communications.

Note that each user can choose her desired tradeoff between privacy protection and overhead by adjusting the countermeasure parameters. Ideally, the countermeasure software can  ask the user her tolerable padding/delay overhead (or her target FP/FN for the adversary), and then will choose the best countermeasure parameters for the user.
For instance, based on Figure~\ref{fig:deepcorr-delay},  assuming that a real-world adversary can tolerate a FP of $10^{-3}$,  if the user states that she intends to keep the adversary's TP below 0.3, the countermeasure software will delay packets with an average of 1s.

\section{Conclusions}
In this paper, we showed  how  popular IM applications leak sensitive information about their clients to adversaries who merely monitor  encrypted traffic.
Specifically, we devised traffic analysis attacks that enable an adversary to identify the administrators and  members of target IM channels with practically high accuracies.
We demonstrated the practicality of  our  attacks through extensive experiments on real-world IM systems.
We believe that our study presents  a significant, real-world threat to the users of such services
given the escalating attempts by oppressive governments in  cracking down on social media.

We  also investigated  the use of standard countermeasures against our attacks and demonstrated their practical feasibility  at the cost of  communication overhead and  increased IM latency.
We designed and implemented   an open-source, publicly available  countermeasure system, \impr, which works  for  major IM services with no need to  support from  the IM providers.
While \impr may be used as an ad hoc, short-term  countermeasure by IM users, we believe that to achieve the  best usability and user adoption, effective countermeasures should be deployed by IM providers (i.e., through integrating traffic obfuscation techniques into their software).
We hope that our study will urge IM providers to take action.

\section*{Acknowledgements}
We would like to thank  Thomas Schneider and Christian Weinert for shepherding our paper, and anonymous reviewers for their constructive feedback.
The work has been supported by the NSF grant CNS-1564067 and by DARPA and NIWC under contract N66001-15-C-4067. The U.S.\ Government is authorized to reproduce and distribute reprints for Governmental purposes notwithstanding any copyright notation thereon. The views, opinions, and/or findings expressed are those of the author(s) and should not be interpreted as representing the official views or policies of the Department of Defense or the U.S.\ Government.

%

\bibliographystyle{IEEEtranS}
\bibliography{ref}

\begin{appendix}

\subsection{Countermeasures Against Shape-Based Detection}\label{cm-cosine}
\begin{figure*}[!htb]
\minipage{0.3\textwidth}
		\includegraphics[width = \linewidth, trim = {1.2cm 1.2cm 1.2cm 1.2cm}]{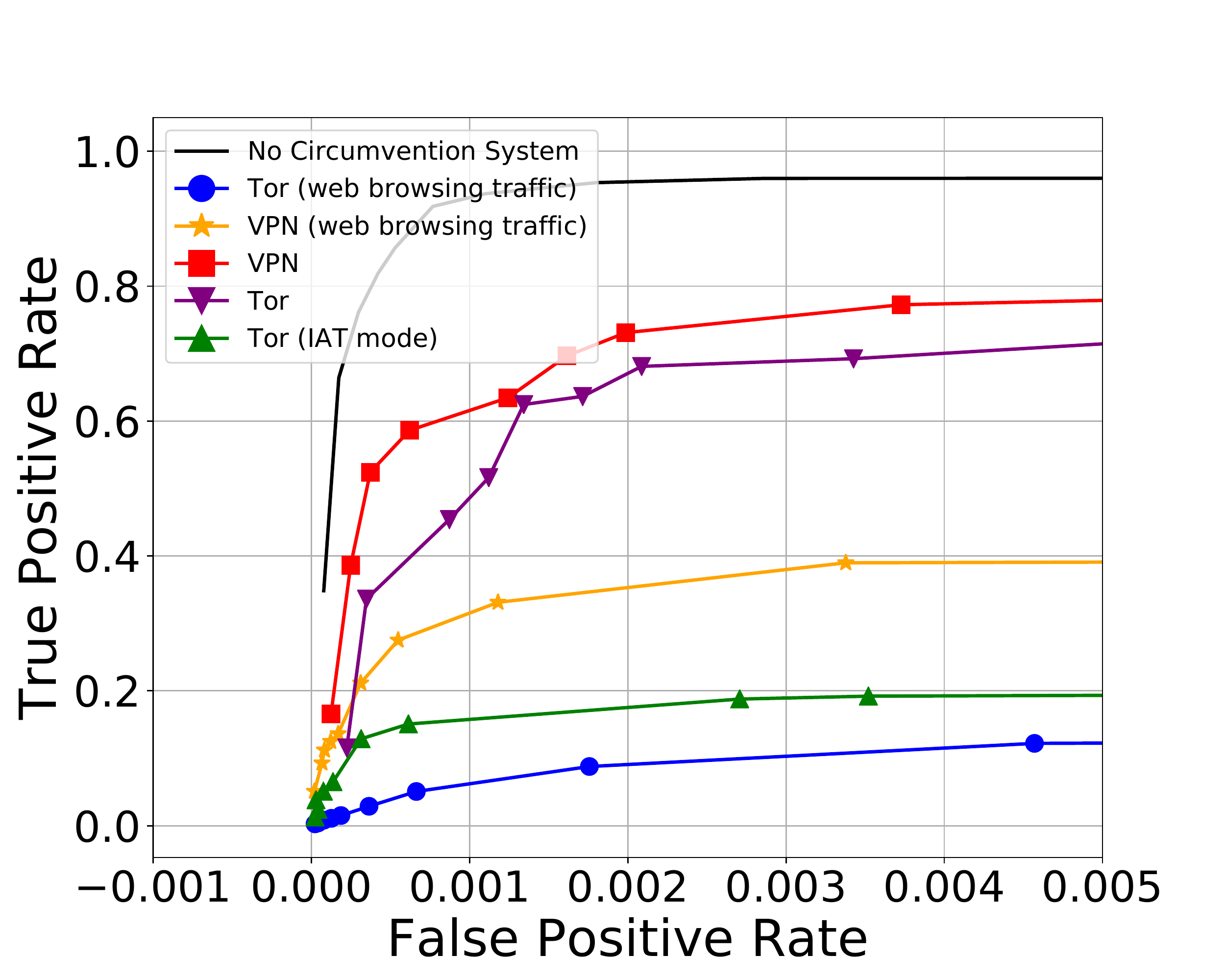}
		\caption{The impact of various countermeasures on the performance of the shape-based detector (15 minutes of observed traffic).}
		\label{fig:roc-cm-cosine}
\endminipage\hfill
\minipage{0.3\textwidth}
		\includegraphics[width = \linewidth, trim = {1.2cm 1cm 1.2cm 1.2cm}]{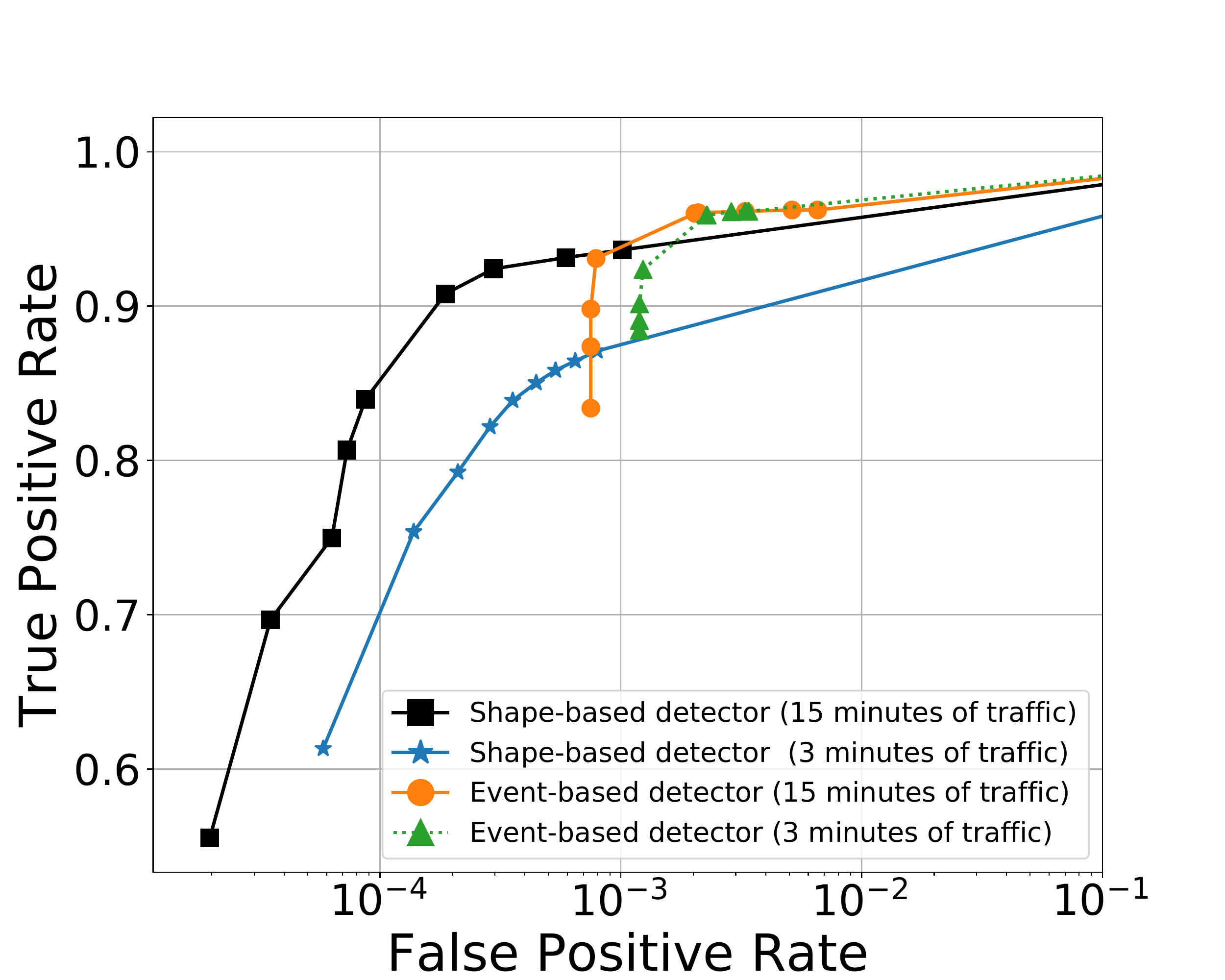}
		\caption{Performance of the shape-based and event-based detectors over 10,000 synthetic IM connections (15 minutes of observed traffic).
		}
		\label{fig:roc-synth}
\endminipage\hfill
\minipage{0.3\textwidth}
\centering
		\includegraphics[width = \linewidth, trim = {1.2cm 1cm 1.2cm 1.2cm}]{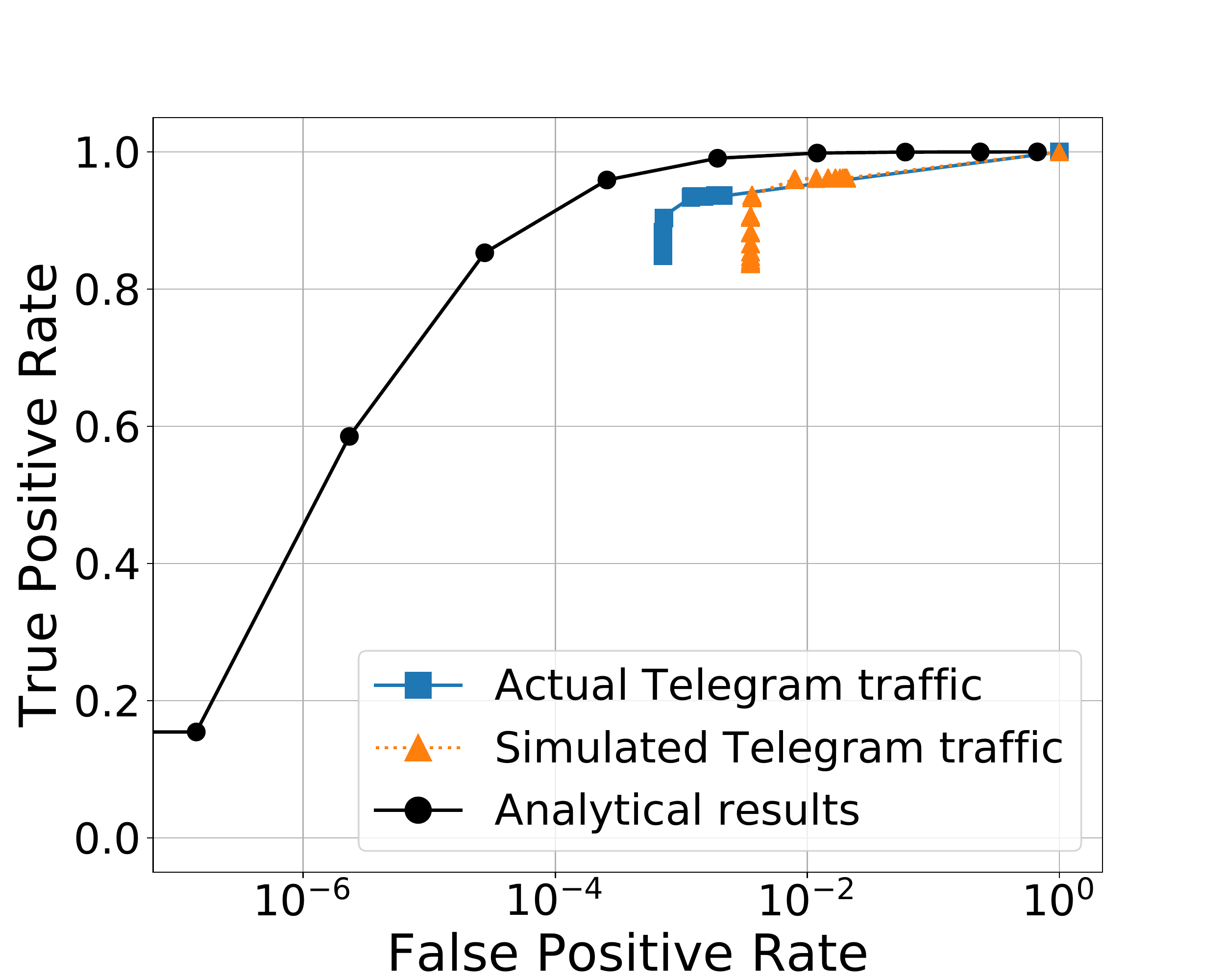}
		\caption{Comparing the analytical upper bounds of the event-based detector (Section~\ref{event-based}) with empirical results (for 15 minutes of traffic).
		}
		\label{fig:fp-fn}
\endminipage
\end{figure*}

Figure~\ref{fig:roc-cm-cosine} shows the performance of the shape-based detector while tunneling traffic through circumvention systems. Experiment setting for Tor, VPN, and background traffic has exactly the same settings as in Section~\ref{tunnel}. However, results show that circumvention systems are more impactful on the shape-based detector. Again, only passing traffic through a VPN or Tor does not effect the performance of the detector significantly, and shape-based detector has about an $80\%$ true positive rate while false positive rate is $0.003$. Similar to Section~\ref{tunnel}, adding web browsing background traffic or using Tor pluggable transports can impact the performance of the shape-based detection.


\subsection{Simulations Using Synthetic Traffic}\label{appendix:simulation}
As  mentioned earlier, major \im services limit the number of channels a client can join;
this limits the reliability of our in-the-wild evaluations.
To address, we generate synthetic \im channels to evaluate our detectors on a much larger number of synthetically created  IM channels.

\paragraphb{Creating Synthetic \im Communications:}
We use  Algorithm~\ref{syn-alg} discussed in Section~\ref{synthesize} to create synthetic \im events (i.e., messages).
To convert these \im events into \im traffic, we need to simulate the impact of network conditions and other perturbations.
Specifically, we apply the effect of network latency according to a Laplacian distribution (see section~\ref{find-model}).
Suppose that $t^{(C)} = \{t^{(C)}_{1},\ldots,t^{(C)}_{n}\}$ and $s^{(C)} = \{s^{(C)}_{1},\ldots,s^{(C)}_{n}\}$ are the vectors of timings and sizes of messages in a synthetic channel, and let $d = \{d_{1},\ldots,d_{n}\}$ be the vector of latencies generated according to a Laplacian distribution. We derive the vector of timings and sizes of the target user flow as follows:
\[	\begin{cases}
	t_i^{(U)}=t_i^{(C)}-d_i, 1 \leq i \leq n  \\
	s_i^{(U)}=s_i^{(C)},  1 \leq i \leq n
	\end{cases}\]
Note that we assume that sizes remain the same in transit.
We also simulate the impact of burst extraction noise.  Assume the user's bandwidth is $bw$.
Let $t_i^{(l)} = \frac{s^{(U)}_{i}}{bw}$ be the time it takes the target user to send the i\textit{th} message. We merge two consecutive messages in the target user's traffic if $t^{(U)}_{i+1} - t^{(U)}_{i} - t^{(l)}_{i} < t_{e}$.


\paragraphb{Comparing our Two Attacks:}
We  apply our shape-based and event-based detectors on the synthetically-generated  \im traffic.
Specifically, we create $10,000$ synthetic \im communications (in contrast to $500$ in the real-world experiments). The channels are divided into five rate buckets.
To evaluate false positives,  we cross-correlate every \im client with all the $10,000-1$ connects in her rate bucket.
Figure~\ref{fig:roc-synth} shows the ROC curve of the shape-based and event-based detectors over synthetic channels.
We can see that, similar to the in-the-wild experiments, the shape-based detector achieves a higher accuracy for smaller false positive rates.

\paragraphb{Comparing To Analytical Bounds}
We also compare our empirical results with the analytical upper bounds of Section~\ref{event-based}.
Figure~\ref{fig:fp-fn} shows the ROC curve of analytical results and experiments using 15 minutes of traffic.
As expected,  for a fixed false positive rate, the analytical results (formula~\ref{fp} and \ref{fn}) upper bound the real-world true positive rate of our event-based
detector for both actual and simulated \tg traffic. Our analytical bound is particularly more useful for smaller values of FP: performing credible experiments for small FPs require a significantly large number of intercepted IM connections which is impractical to capture in experiments.


%

\subsection{Overhead of Padding Through Dummy Events}\label{padding}
{
\begin{table}[!ht]
\centering
\caption{Bandwidth overheads of padding through dummy events  for different values of $p_{padding}$.}
\resizebox{0.7\linewidth}{!}{
\begin{tabular}{ |c|c| }
	\hline
	$p_{padding}$ & \textbf{Bandwidth Overhead} \\
	\hline
	0.0001 & 7\% \\
	\hline
	0.0005 & 34\% \\
	\hline
	0.001 & 67\% \\
	\hline
\end{tabular}}
\label{table:padding}
\end{table}
}
Here, we calculate the bandwidth overhead of  adding dummy events as introduced in Section~\ref{obfs}. We  assume that the number of periods in which there is a dummy event follows a Binomial distribution with parameters $p_{padding}$ and the size of the observed flow.
Therefore,
the average number of dummy events in each flow will  follow the mean of the Binomial distribution, which is equal to $p_{padding}\times \ell$, where $\ell$ is the length of observed flows in seconds.
The size of each dummy event is sampled from our collection of IM image messages, which has an average size of  $90Kb$.
Therefore, we can evaluate the bandwidth overhead of dummy events by dividing the average volume of dummy events over the volume of actual IM messages in long traffic observations.
This is shown in Table~\ref{padding}  for different values of $p_{padding}$.

\subsection{Rate-Based Transition Matrices}
\label{tr}

The following is the empirical transition probability matrix of the Markov model we use to model IM message sizes (in Section~\ref{find-model}) for the aggregation of all channels:

\[
	P=
	\left[ {\begin{array}{ccccc}
	0.40 & 0.47 & 0.10 & 0.01 & 0.02 \\
	0.29 & 0.53 & 0.11 & 0.02 & 0.05 \\
	0.19 & 0.36 & 0.40 & 0.02 & 0.03 \\
	0.17 & 0.59 & 0.13 & 0.09 & 0.02 \\
	0.14 & 0.40 & 0.10 & 0.01 & 0.35 \\
	\end{array} } \right]
\]

The following are the transition matrices for groups of channels  with different average daily message rates of 2.31, 7.68, 18.34, 39.47, and 130.57, respectively. We see that the models change slightly for different types of channels.

{\tiny
\[
	P_{1}=
	\begin{bmatrix}
	0.48 & 0.41 & 0.07 & 0.00 & 0.04 \\
	0.28 & 0.52 & 0.11 & 0.01 & 0.08 \\
	0.12 & 0.32 & 0.49 & 0.00 & 0.07 \\
	0.14 & 0.43 & 0.14 & 0.00 & 0.29 \\
	0.13 & 0.44 & 0.06 & 0.00 & 0.38
	\end{bmatrix}
	P_{2}=
	\begin{bmatrix}
	0.55 & 0.28 & 0.10 & 0.02 & 0.05 \\
	0.18 & 0.59 & 0.11 & 0.01 & 0.12 \\
	0.13 & 0.35 & 0.45 & 0.01 & 0.07 \\
	0.17 & 0.36 & 0.14 & 0.33 & 0.00 \\
	0.19 & 0.34 & 0.10 & 0.03 & 0.33
	\end{bmatrix}
\]
%
\[
	P_{3}=
	\begin{bmatrix}
	0.45 & 0.38 & 0.12 & 0.02 & 0.04 \\
	0.22 & 0.55 & 0.13 & 0.03 & 0.07 \\
	0.15 & 0.35 & 0.42 & 0.04 & 0.05 \\
	0.15 & 0.54 & 0.20 & 0.51 & 0.06 \\
	0.13 & 0.31 & 0.10 & 0.03 & 0.43
	\end{bmatrix}
	P_{4}=
	\begin{bmatrix}
	0.38 & 0.44 & 0.14 & 0.02 & 0.02 \\
	0.24 & 0.50 & 0.15 & 0.03 & 0.09 \\
	0.17 & 0.35 & 0.43 & 0.03 & 0.03 \\
	0.20 & 0.55 & 0.15 & 0.09 & 0.01 \\
	0.09 & 0.46 & 0.11 & 0.01 & 0.33
	\end{bmatrix}
\]
%
\[
	P_{5}=
	\begin{bmatrix}
	0.40 & 0.48 & 0.09 & 0.01 & 0.01 \\
	0.32 & 0.53 & 0.10 & 0.02 & 0.04 \\
	0.21 & 0.37 & 0.39 & 0.02 & 0.02 \\
	0.16 & 0.63 & 0.11 & 0.08 & 0.01 \\
	0.16 & 0.41 & 0.10 & 0.01 & 0.32
	\end{bmatrix}
\]
}
%
%

\end{appendix}

\end{document}